\documentclass[aps,pre,showpacs,showkeys,onecolumn,superscriptaddress,amsmath,amssymb]{revtex4}
\usepackage[dvips]{color}
\usepackage[dvips]{graphicx}

\newcommand{\ie}{\emph{i.e.}, }
\newcommand{\TT}{\mathrm{T}}
\newcommand{\Tr}{\mathrm{Tr}}

\newcommand{\re}{\mathrm{Re}}
\newcommand{\im}{\mathrm{Im}}

\newcommand{\diag}{\mathrm{diag}}
\newcommand{\la}{\left\langle}
\newcommand{\ra}{\right\rangle}
\newcommand{\MAq}{\mathrm{MA} ( q )}
\newcommand{\VMAq}{\mathrm{VMA} ( q )}
\newcommand{\VMAqOne}{\mathrm{VMA} ( q_{1} )}
\newcommand{\VMAqTwo}{\mathrm{VMA} ( q_{2} )}

\newcommand{\ARq}{\mathrm{AR} ( q )}
\newcommand{\VARq}{\mathrm{VAR} ( q )}

\newcommand{\VARMA}{\mathrm{VARMA} ( q_{1} , q_{2} )}
\newcommand{\VARMAOneOne}{\mathrm{VARMA} ( 1 , 1 )}
\newcommand{\BN}{\mathbf{N}}

\begin{document}

\title{A Random Matrix Approach to Dynamic Factors in macroeconomic data}
\author{Ma{\l}gorzata Snarska}
\email{snarska@th.if.uj.edu.pl}
\affiliation{Marian Smoluchowski Institute of Physics and Mark Kac Complex Systems Research Centre, Jagiellonian University, Reymonta 4, 30--059 Krak\'{o}w, Poland}
\affiliation{Cracow University of Economics, Department of Econometrics and Operations Research, Rakowicka 27, 31--510 Krak\'{o}w, Poland}

\date{\today}

\setlength{\parindent}{4ex}
\setlength{\parskip}{1.5ex plus 0.5ex minus 0.2ex}
\topmargin=0cm

\begin{abstract}
We show how random matrix theory can be applied to develop new algorithms to extract dynamic factors from macroeconomic time series. In particular, we consider a limit where the number of random variables $N$ and the number of consecutive time measurements $T$ are large but the ratio $N / T$ is fixed. In this regime the underlying random matrices are asymptotically equivalent to Free Random Variables (FRV).Application of these methods for macroeconomic indicators for Poland economy is also presented.
\end{abstract}


\maketitle


\section{Introduction}
\label{s:Introduction}

Macroeconomic time series are characterized by two main features:comovements and cyclic phases of expansion and depression. Behind the econometric literature, especially on business cycle and VAR modeling there is an idea, that essential characteristics of macroeconomic time series are adequately captured by a small number of nearly independent factors. \\
On the other hand many empirical studies suggest to analyze many variables in order to understand the microeconomic mechanisms behind the fluctuations, where both $N$- the number of variables and $T$ the length of a time series are typically large, and typically of the same order.
Surprisingly most of the models developed in this field of science can precisely describe only static properties
of the correlation structure.
There is a strong need for models and algorithms, that can accurately capture the above described features of
macroeconomic time series,are dynamic and allow for a reacher spatio -- temporal structure.
In this paper we present new algorithms, based on Random Matrix Theory that countenance extracting potentially useful factors in macroeconomic time series. In the first section we uncover external temporal correlation structure, using the assumption that the characteristic of each individual time series is meticulously depicted by the $\VARMA$ structure with the same parameters. The second method encompasses exposition of internal correlation structure between two different sets of variables, after the external correlations are properly reduced.


\section{Unraveling Lagged External Temporal Correlations from VARMA(p,q)}
\label{s:DoublyCorrelatedWishartEnsemblesAndFreeRandomVariables}

Finite order vector autoregressive moving average models (VARMA) motivated by Wold decomposition theorem
\cite{Wold1938} as an appriopriate multivariate setting for studying the dynamics of stationary time series.
Vector autoregressive (VAR) models are cornerstones in contemporary macroeconomics, being a part of  an approach called the ``dynamic stochastic general equilibrium''(DSGE), which is superseding traditional large--scale macroeconometric forecasting methodologies~\cite{Sims1980}. The motivation behind them is based on the assertion that more recent values of a variable are more likely to contain useful information about its future movements than the older ones. On the other hand, a standard tool in multivariate time series analysis is vector moving average (VMA) models, which is really a linear regression of the present value of the time series w.r.t. the past values of a white noise. A broader class of stochastic processes used in macroeconomics comprises both these kinds together in the form of vector autoregressive moving average (VARMA) models. These methodologies can capture certain spatial and temporal structures of multidimensional variables which are often neglected in practice; including them not only results in more accurate estimation, but also leads to models which are more interpretable.


\subsection{Correlated Gaussian Random Variables}
\label{sss:CorrelatedGaussianRandomVariables}


We will consider a situation of $N$ time--dependent random variables which are measured at $T$ consecutive time moments (separated by some time interval $\delta t$); let \smash{$Y_{i a}$} be the value od the $i$--th ($i = 1 , \ldots , N$) random number at the $a$--th time moment ($a = 1 , \ldots , T$); together, they make up a rectangular $N \times T$ matrix $\mathbf{Y}$. Firstly we will assume, each \smash{$Y_{i a}$} is supposed to be drawn from a Gaussian probability distribution, and that they have mean values zero, \smash{$\langle Y_{i a} \rangle = 0$}. A set of correlated zero--mean Gaussian numbers is fully characterized by the two--point covariance function \label{eq:CovarianceFunctionDefinition} , \smash{$\mathcal{C}_{i a , j b} \equiv \langle Y_{i a} Y_{j b} \rangle$} if the underlying stochastic process generating these numbers is stationary. The stationarity condition for $\VARMA$ stochastic processes implies certain restrictions on their parameters; for details, we refer to~\cite{Lutkepohl2005}. We will restrict our attention to an even narrower class where the cross--correlations between different variables and the auto--correlations between different time moments are factorized, \ie
\begin{equation}\label{eq:CovarianceFunctionFactorization}
\la Y_{i a} Y_{j b} \ra = C_{i j} A_{a b} .
\end{equation}


\subsubsection{Estimating External Cross -- Covariances}
\label{sss:EstimatingEqualTimeCrossCovariances}

 The realized cross--covariance between degrees $i$ and $j$ at the same time $a$ is \smash{$Y_{i a} Y_{j a}$}, the simplest method to estimate the today's cross--covariance \smash{$c_{i j}$} is to compute the time average (named ''Pearson estimator''),
\begin{equation}\label{eq:EstimatorcDefinition}
c_{i j} \equiv \frac{1}{T} \sum_{a = 1}^{T} Y_{i a} Y_{j a} , \qquad \textrm{\ie} \qquad \mathbf{c} = \frac{1}{T} \mathbf{Y} \mathbf{Y}^{\TT} = \frac{1}{T} \sqrt{\mathbf{C}} \widetilde{\mathbf{Y}} \mathbf{A} \widetilde{\mathbf{Y}}^{\TT} \sqrt{\mathbf{C}} .
\end{equation}
The random matrix $\mathbf{c}$ is called a ``doubly correlated Wishart ensemble''~\cite{Wishart1928}.

It is obvious, that our estimator will reflect the true covariances only to a certain degree, with a superimposed broadening due to the finiteness of the time series \ie estimation accuracy will depend on the ``rectangularity ratio,''
\begin{equation}\label{eq:NoiseToSignalRatio}
r \equiv \frac{N}{T} ;
\end{equation}
the closer $r$ to zero, the more truthful the estimate. Furthermore we will consider the situation, where
\begin{equation}\label{eq:ThermodynamicalLimit}
N \to \infty , \qquad T \to \infty , \qquad \textrm{such that} \qquad r = \textrm{fixed} .
\end{equation}


\subsection{Free Random Variables Calculus Crash Course}
\label{ss:AShortIntroductionToTheFreeRandomVariablesCalculusTheMultiplicationAlgorithm}


\subsubsection{The $M$--Transform and the Spectral Density}
\label{sss:TheMTransformAndTheSpectralDensity}

Let's consider (real symmetric $K \times K$) random matrix $\mathbf{H}$ . The  average of its eigenvalues \smash{$\lambda_{1} , \ldots , \lambda_{K}$}is concisely encoded in the ``mean spectral density,''
\begin{equation}\label{eq:MeanSpectralDensityDefinition}
\rho_{\mathbf{H}} ( \lambda ) \equiv \frac{1}{K} \sum_{i = 1}^{K} \la \delta \left( \lambda - \lambda_{i} \right) \ra = \frac{1}{K} \la \Tr \left( \lambda \mathbf{1}_{K} - \mathbf{H} \right) \ra .
\end{equation}
On the practical side, it is more convenient to work with either of the two equivalent objects,
\begin{equation}\label{eq:GreenFunctionAndMTransformDefinition}
G_{\mathbf{H}} ( z ) \equiv \frac{1}{K} \la \Tr \frac{1}{z \mathbf{1}_{K} - \mathbf{H}} \ra , \qquad \textrm{or} \qquad M_{\mathbf{H}} ( z ) \equiv z G_{\mathbf{H}} ( z ) - 1 ,
\end{equation}
referred to as the ``Green's function'' (or the ``resolvent'') and the ``$M$--transform'' of $\mathbf{H}$ (``moments' generating function,''). The latter one serves better in the case of multiplication of random matrices.


\subsubsection{The $N$--Transform and Free Random Variables}
\label{sss:TheNTransformAndFreeRandomVariables}

The doubly correlated Wishart ensemble $\mathbf{c}$ (\ref{eq:EstimatorcDefinition}) may be viewed as a product of several random and non--random matrices. The general problem of multiplying random variables in classical probability theory can be effectively handled in the special situation when the random terms are independent: then, the exponential map reduces it to the addition problem of independent random numbers, solved by considering the logarithm of the characteristic functions of the respective PDFs, which proves to be additive. In matrix probability theory due to D.~Voiculescu and coworkers and R.~Speicher~\cite{VoiculescuDykemaNica1992,Speicher1994}, we have a parallel  commutative construction in the noncommutative world (c.f. table \ref{tab:FRVanalog}). It starts with the notion of ``freeness,'' which basically comprises probabilistic independence together with a lack of any directional correlation between two random matrices. This nontrivial new property happens to be the right extension of classical independence, as it allows for an efficient algorithm of multiplying free random variables (FRV), which we state below:
\begin{description}
\item[Step 1:] Suppose we have two random matrices, \smash{$\mathbf{H}_{1}$} and \smash{$\mathbf{H}_{2}$}, mutually free. Their spectral properties are best wrought into the $M$--transforms (\ref{eq:GreenFunctionAndMTransformDefinition}), \smash{$M_{\mathbf{H}_{1}} ( z )$} and \smash{$M_{\mathbf{H}_{2}} ( z )$}.
\item[Step 2:] The critical maneuver is to turn attention to the functional inverses of these $M$--transforms, the so--called ``$N$--transforms,''
    \begin{equation}\label{eq:NTransformDefinition}
    M_{\mathbf{H}} \left( N_{\mathbf{H}} ( z ) \right) = N_{\mathbf{H}} \left( M_{\mathbf{H}} ( z ) \right) = z .
    \end{equation}
\item[Step 3:] The $N$--transforms submit to a very straightforward rule upon multiplying free random matrices (the ``FRV multiplication law''),
    \begin{equation}\label{eq:FRVMultiplicationLaw}
    N_{\mathbf{H}_{1} \mathbf{H}_{2}} ( z ) = \frac{z}{1 + z} N_{\mathbf{H}_{1}} ( z ) N_{\mathbf{H}_{2}} ( z ) , \qquad \textrm{for free \smash{$\mathbf{H}_{1}$}, \smash{$\mathbf{H}_{2}$}.}
    \end{equation}
\item[Step 4:] Finally, it remains to functionally invert the resulting $N$--transform\\
$N_{\mathbf{H}_{1} \mathbf{H}_{2}} ( z )$ to gain the $M$--transform of the product, $M_{\mathbf{H}_{1} \mathbf{H}_{2}} ( z )$.
\end{description}

\begin{table}[!t]
\begin{tabular}{|cc|}\hline
\textbf{Classical Probability}& \textbf{Noncommutative probability (FRV)}\\
$x$ - random variable, $p(x)$ & $H$ - random matrix, $P(H)$ \\
pdf & spectral density $\varrho(\lambda)d\lambda$ \\
characteristic function $g_x(z)\equiv\la e^{izx}\ra$ & Green's function $G_H(z)=\frac{1}{N}\la\Tr\frac{1}{z\cdot\mathbf{1}- H}\ra$ \\& or  M - transform $M(z)=zG_H(z)-1$   \\
independence & freeness\\
Addition of independent r.v.: & Addition of f.r.v.\\
The logarithm of the characteristic function,& The Blue's function\\
$r_x(z)\equiv\log g_x(z)$, is additive,&
$
G_H (B_H(z)) = B_H (G_H(z)) = z$, is additive,\\
$
r_{x_1+x_2} (z) = r_{x_1} (z) + r_{x_2} (z)
$ &
$
B_{H_1+H_2} (z) = B_{H_1} (z) + B_{H_2} (z) -\frac1z
$\\& \\
Multiplication of independent r.v.:&Multiplication of free r.v.:\\
 Reduced to the
addition problem & The N - transform,\\
via the exponential map, owing to  & $M_H (N_H(z)) = N_H (M_H(z)) = z$,\\
$e^{x_1}e^{x_2} = e^{x_1+x_2}$& is multiplicative\\
&$
N_{H_1H_2} (z) =\frac{z}{1+z}N_{H_1} (z)N_{H_2} (z)$\\ &\\ \hline
\end{tabular}
  \caption{Parallel between Classical and Noncommutative probability}\label{tab:FRVanalog}
\end{table}


\subsubsection{Extracting External Correlations}
\label{s:VARMAFromFreeRandomVariables}

The innate potential of the FRV multiplication algorithm (\ref{eq:FRVMultiplicationLaw}) is surely revealed when inspecting the doubly correlated Wishart random matrix \smash{$\mathbf{c} = ( 1 / T ) \sqrt{\mathbf{C}} \widetilde{\mathbf{Y}} \mathbf{A} \widetilde{\mathbf{Y}}^{\TT} \sqrt{\mathbf{C}}$} (\ref{eq:EstimatorcDefinition}). This has been done in detail in~\cite{BurdaJaroszJurkiewiczNowakPappZahed2009,BurdaJaroszNowakSnarska2010}, so we will only accentuate the main results here, referring the reader to the original papers for an exact explanation.
The idea is that one uses twice the cyclic property of the trace (which permits cyclic shifts in the order of the terms), and twice the FRV multiplication law (\ref{eq:FRVMultiplicationLaw}) (to break the $N$--transforms of products of matrices down to their constituents), in order to reduce the problem to solving the uncorrelated Wishart ensemble \smash{$( 1 / T ) \widetilde{\mathbf{Y}}^{\TT} \widetilde{\mathbf{Y}}$}. This last model is further simplified, again by the cyclic property and the FRV multiplication rule applied once, to the standard $\mathbf{GOE}$ random matrix squared (and the projector \smash{$\mathbf{P} \equiv \diag ( \mathbf{1}_{N} , \mathbf{0}_{T - N} )$}, designed to chip the rectangle \smash{$\widetilde{\mathbf{Y}}$} off the square $\mathbf{GOE}$), whose properties are firmly established. Let us sketch the derivation,
$$
N_{\mathbf{c}} ( z ) \stackrel{\substack{\mathrm{cyclic}\\\downarrow}}{=} N_{\frac{1}{T} \widetilde{\mathbf{Y}} \mathbf{A} \widetilde{\mathbf{Y}}^{\TT} \mathbf{C}} ( z ) \stackrel{\substack{\mathrm{FRV}\\\downarrow}}{=} \frac{z}{1 + z} N_{\frac{1}{T} \widetilde{\mathbf{Y}} \mathbf{A} \widetilde{\mathbf{Y}}^{\TT}} ( z ) N_{\mathbf{C}} ( z ) \stackrel{\substack{\mathrm{cyclic}\\\downarrow}}{=}
$$
$$
\stackrel{\substack{\mathrm{cyclic}\\\downarrow}}{=} \frac{z}{1 + z} N_{\frac{1}{T} \widetilde{\mathbf{Y}}^{\TT} \widetilde{\mathbf{Y}} \mathbf{A}} ( r z ) N_{\mathbf{C}} ( z ) \stackrel{\substack{\mathrm{FRV}\\\downarrow}}{=} \frac{z}{1 + z} \frac{r z}{1 + r z} N_{\frac{1}{T} \widetilde{\mathbf{Y}}^{\TT} \widetilde{\mathbf{Y}}} ( r z ) N_{\mathbf{A}} ( r z ) N_{\mathbf{C}} ( z ) =
$$
\begin{equation}\label{eq:DoublyCorrelatedWishartEnsembleTheMainEquation1}
= r z N_{\mathbf{A}} ( r z ) N_{\mathbf{C}} ( z ) .
\end{equation}

This is the basic formula. Since the spectral properties of $\mathbf{c}$ are given by its $M$--transform, \smash{$M \equiv M_{\mathbf{c}} ( z )$}, it is more pedagogical to recast (\ref{eq:DoublyCorrelatedWishartEnsembleTheMainEquation1}) as an equation for the unknown $M$,
\begin{equation}\label{eq:DoublyCorrelatedWishartEnsembleTheMainEquation2}
z = r M N_{\mathbf{A}} ( r M ) N_{\mathbf{C}} ( M ) .
\end{equation}
It provides a means for computing the mean spectral density of a doubly correlated Wishart random matrix once the ``true'' covariance matrices $\mathbf{C}$ and $\mathbf{A}$ are given.

In this communication, only a particular instance of this fundamental formula is applied, namely with an arbitrary auto--covariance matrix $\mathbf{A}$, but with trivial cross--covariances, \smash{$\mathbf{C} = \mathbf{1}_{N}$}. Using that \smash{$N_{\mathbf{1}_{K}} ( z ) = 1 + 1 / z$}, equation (\ref{eq:DoublyCorrelatedWishartEnsembleTheMainEquation2}) thins out to
\begin{equation}\label{eq:DoublyCorrelatedWishartEnsembleTheMainEquation3}
r M = M_{\mathbf{A}} \left( \frac{z}{r ( 1 + M )} \right) ,
\end{equation}
which will be strongly exploited below.

For all this, we must in particular take both $N$ and $T$ large from the start, with their ratio $r \equiv N / T$ fixed (\ref{eq:ThermodynamicalLimit}). More precisely, we stretch the range of the $a$--index from minus to plus infinity. This means that all the finite--size effects (appearing at the ends of the time series) are readily disregarded. In particular, there is no need to care about initial conditions for the processes, and all the recurrence relations are assumed to continue to the infinite past.


\subsection{The \smash{$\VARMA$} Process}
\label{ss:TheVARMAq1q2Process}


\subsubsection{The Definition of \smash{$\VARMA$}}
\label{sss:VARMAq1q2TheDefinition}

One can define stochastic proces called \smash{$\VARMA$} in the following way as a convolution of $\VARq$ and $\VMAq$ processes:
\begin{equation}\label{eq:VARMAq1q2Definition}
Y_{i a} - \sum_{\beta = 1}^{q_{1}} b_{\beta} Y_{i , a - \beta} = \sum_{\alpha = 0}^{q_{2}} a_{\alpha} \epsilon_{i , a - \alpha} .
\end{equation}
To be more specific, we consider a situation when $N$ stochastic variables evolve according to identical independent  \smash{$\VARMA$}(vector autoregressive moving average processes, , which we sample over a time span of $T$ moments. 

$\VMAq$ (vector moving average) process is a simple generalization of the standard univariate weak--stationary moving average $\MAq$. In such a setting, the value \smash{$Y_{i a}$} of the $i$--th ($i = 1 , \ldots , N$) random variable at time moment $a$ ($a = 1 , \ldots , T$) can be expressed as
\begin{equation}\label{eq:VMAqDefinition}
Y_{i a} = \sum_{\alpha = 0}^{q} a_{\alpha} \epsilon_{i , a - \alpha} .
\end{equation}
Here all the \smash{$\epsilon_{i a}$}'s are IID standard (mean zero, variance one) Gaussian random numbers (white noise), \smash{$\langle \epsilon_{i a} \epsilon_{j b} \rangle = \delta_{i j} \delta_{a b}$}. The \smash{$a_{\alpha}$}'s are some $( q + 1 )$ real constants; importantly, they do not depend on the index $i$, which reflects the fact that the processes are identical and independent (no ``spatial'' covariances among the variables). The rank $q$ of the process is a positive integer.

 The $\VARq$ (vector auto--regressive) processes  part is somewhat akin to (\ref{eq:VMAqDefinition}), \ie we consider $N$ decoupled copies of a standard univariate $\ARq$ process,
\begin{equation}\label{eq:VARqDefinition}
Y_{i a} - \sum_{\beta = 1}^{q} b_{\beta} Y_{i , a - \beta} = a_{0} \epsilon_{i a} .
\end{equation}
It is again described by the demeaned and standardized Gaussian white noise \smash{$\epsilon_{i a}$} (which triggers the stochastic evolution), as well as $( q + 1 )$ real constants \smash{$a_{0}$}, \smash{$b_{\beta}$}, with $\beta = 1 , \ldots , q$. As announced before, the time stretches to the past infinity, so no initial condition is necessary. Although at first sight (\ref{eq:VARqDefinition}) may appear to be a more involved recurrence relation for the \smash{$Y_{i a}$}'s, it is actually easily reduced to the $\VMAq$ case: It remains to remark that if one exchanges the \smash{$Y_{i a}$}'s with the \smash{$\epsilon_{i a}$}'s, one precisely arrives at the $\VMAq$ process with the constants \smash{$a^{( 2 )}_{0} \equiv 1 / a_{0}$}, \smash{$a^{( 2 )}_{\beta} \equiv - b_{\beta} / a_{0}$}, $\beta = 1 , \ldots , q$. In other words, the auto--covariance matrix \smash{$\mathbf{A}^{( 3 )}$} of the $\VARq$ process (\ref{eq:VARqDefinition}) is simply the inverse of the auto--covariance matrix \smash{$\mathbf{A}^{( 2 )}$} of the corresponding $\VMAq$ process with the described modification of the parameters,
\begin{equation}\label{eq:VARqFromVMAq}
\mathbf{A}^{( 3 )} = \left( \mathbf{A}^{( 2 )} \right)^{- 1} .
\end{equation}
This inverse exists thanks to the weak stationarity supposition.

The auto--covariance matrix \smash{$\mathbf{A}^{( 5 )}$} of this process is simply the product (in any order) of the auto--covariance matrices of the VAR and VMA pieces; more precisely,
\begin{equation}\label{eq:VARMAq1q2FromVARq1VMAq2}
\mathbf{A}^{( 5 )} = \left( \mathbf{A}^{( 4 )} \right)^{- 1} \mathbf{A}^{( 1 )} ,
\end{equation}
where \smash{$\mathbf{A}^{( 1 )}$} corresponds to the generic \smash{$\VMAqTwo$} model, while \smash{$\mathbf{A}^{( 4 )}$} denotes the auto--covariance matrix of \smash{$\VMAqOne$} with a slightly different modification of the parameters compared to the previously used, namely \smash{$a^{( 4 )}_{0} \equiv 1$}, \smash{$a^{( 4 )}_{\beta} \equiv - b_{\beta}$}, for \smash{$\beta = 1 , \ldots , q_{1}$}.

The $M$--transform of \smash{$\mathbf{A}^{( 5 )}$} can consequently be derived from the general formulas \cite{BurdaJaroszNowakSnarska2010}. We will evaluate here the pertinent integral only for the simplest $\VARMAOneOne$ process, even though an arbitrary case may be handled by the technique of residues,
\begin{equation}\label{eq:VARMAOneOneMomentsGeneratingFunctionForAFive}
M_{\mathbf{A}^{( 5 )}} ( z ) = \frac{1}{a_{0} a_{1} + b_{1} z} \left( - a_{0} a_{1} + \frac{z \left( a_{0} a_{1} + \left( a_{0}^{2} + a_{1}^{2} \right) b_{1} + a_{0} a_{1} b_{1}^{2} \right)}{\sqrt{\left( 1 - b_{1} \right)^{2} z - \left( a_{0} + a_{1} \right)^{2}} \sqrt{\left( 1 + b_{1} \right)^{2} z - \left( a_{0} - a_{1} \right)^{2}}} \right) .
\end{equation}


\subsection{Example from Macroeconomic Data}
We apply the method described above to Polish macroeconomic data. The motivation behind is twofold. First of all
economic theory rarely has any sharp implications about the short-run dynamics of economic variables (so called scarcity of economic time series). Secondly in these very rare situations, where theoretical models include a dynamic adjustment equation, one has to work hard to exclude the moving average terms from appearing in the implied dynamics of the variables of interest.
\subsubsection{An Application to Macroeconomic Data}
\label{sss:VARMAq1q2AnApplicationToMacroeconomicData}
\begin{figure}[!h]
\begin{center}
\includegraphics[width=0.45\textwidth]{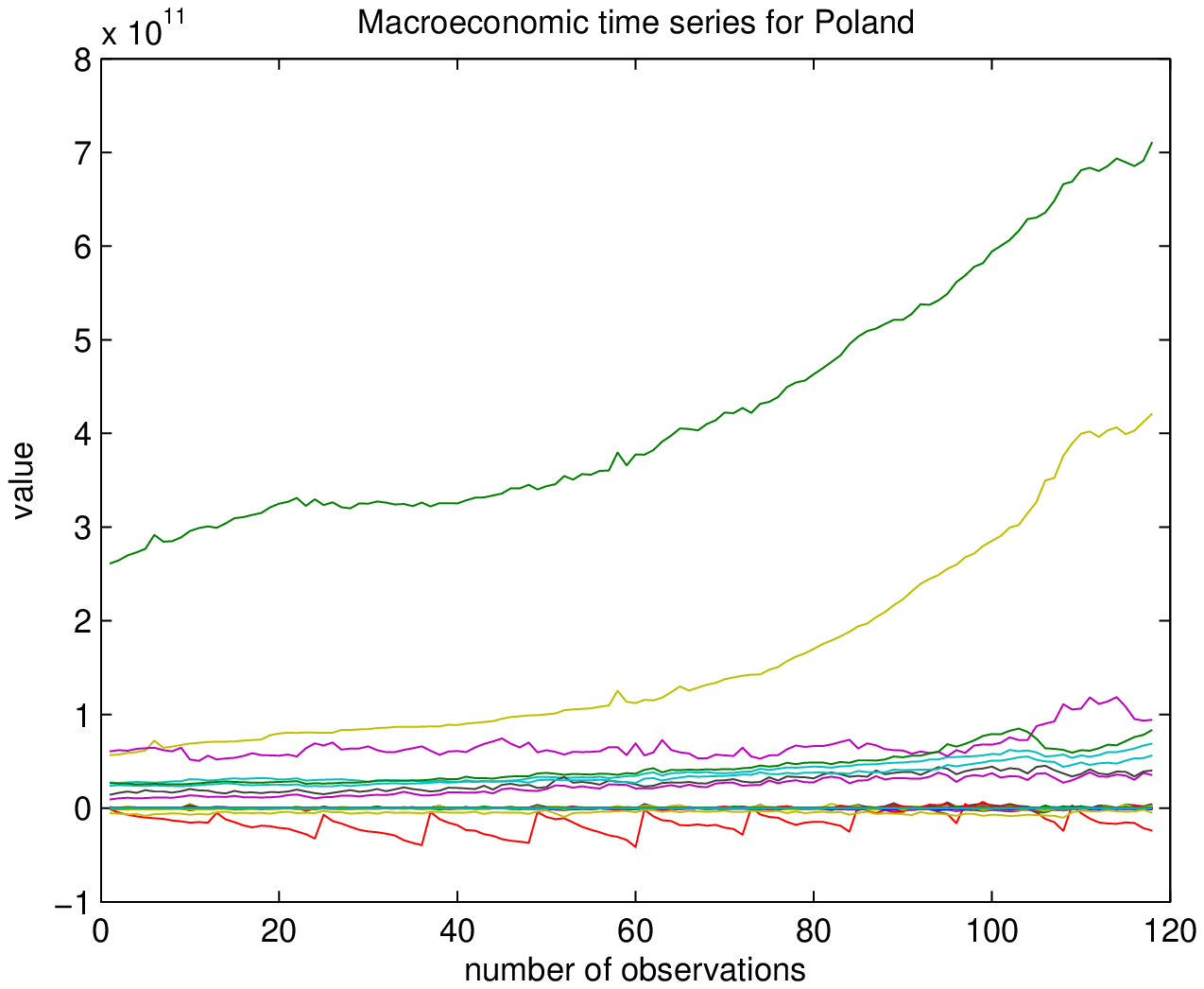}
\includegraphics[width=0.45\textwidth]{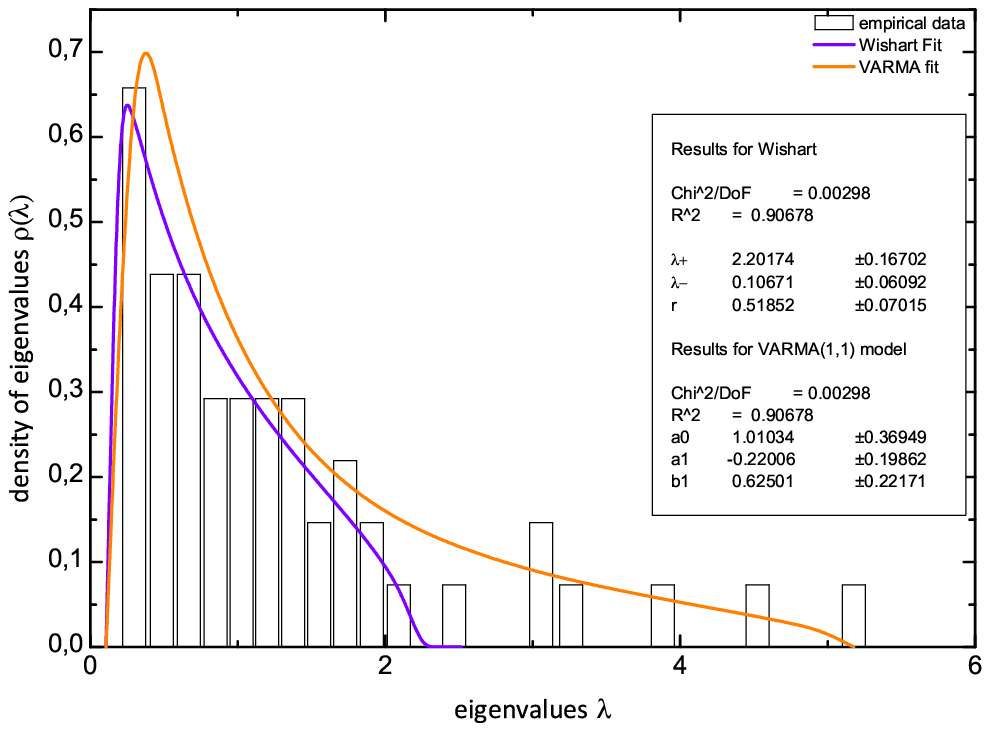}
\caption[Spectral density for real macroeconomic data]{LEFT: The original $N = 52$ time series of length $T = 118$; they are non--stationary, with the seasonal components.\\RIGHT: The histogram of the mean spectral density \smash{$\rho_{\mathbf{c}} ( \lambda )$} (the solid black line) compared to the theoretical result obtained by numerically solving the resulting sixth--order equation (\ref{eq:VARMAOneOneMomentsGeneratingFunctionForAFive}) (the solid orange line) and the ''Wishart-fit'' (purple line).}
\label{fig:VARMAOneOneTheoryPlusData}
\end{center}
\end{figure}

Let us pursue further the above analysis of the $\VARMAOneOne$ model on a concrete example of real data. Economists naturally think of co-movement in economic time series as arising largely
from relatively few key economic factors like productivity, monetary policy and so forth. Classical way of representing this notion is in terms
of statistical factor model, for which we allow the limited spatio--temporal dependence expressed via our $\VARMAOneOne$ model. Two empirical questions are addressed in this section:
\begin{enumerate}
\item How the empirical data behave? Does the eigenvalues represent similar structure to financial market data? In other words, does the macroeconomic data represent collective response to the shock expressed in term of $\VARMAOneOne$ model or are there any apparent outliers.
\item Should then the forecasts be constructed using small factor models or are there any non-zero, but perhaps small coefficients. If so, then the large scale model framework is appropriate.
\end{enumerate}
We investigate $N = 52$ various macroeconomic time series for Poland of length $T = 118$.  They have been selected on a monthly basis in such a manner so as to cover most of the main sectors of the Polish economy, \ie the money market, domestic and foreign trade, labor market, balance of payments, inflation in different sectors, \emph{etc.} The time series were taken directly from the Reuters\copyright 3000Xtra database. Although longer time series for Poland are accessible, we restrict ourselves to the last ten years in order to avoid the effects of structural change. We assume that each economic variable is affected by the same shock (\ie the ``global market shock'') of an $\mathrm{ARMA} ( 1 , 1 )$ type with unknown parameters, which we are to estimate; the AR part implies that the shock dies quite quickly, while the MA part is responsible for the persistency of the shock. To preserve the proper VARMA representation, the original time series were transformed using one of the following methods:
\begin{itemize}
\item First, many of the series are seasonally
adjusted by the reporting agency.
\item Second, the data were transformed to eliminate trends
and obvious nonstationarities. For real variables, this typically involved transformation to
growth rates (the first difference of logarithms), and for prices this involved
transformation to changes in growth rates (the second difference of logarithms).
\item Interest
rates were transformed to first differences.
\item Finally, some of the series
contained a few large outliers associated with events like labor disputes, other extreme
events, or with data problems of various sorts. These outliers were identified as
observations that differed from the sample median by more than $6$ times the sample
interquartile range, and these observations were dropped from the analysis.
\end{itemize}
In fig.~\ref{fig:VARMAOneOneTheoryPlusData} we plot these time series (LEFT), and we make (RIGHT) a histogram of the mean spectral density \smash{$\rho_{\mathbf{c}} ( \lambda )$}, which we compare to a theoretical prediction solving the FRV equation  for (\ref{eq:VARMAOneOneMomentsGeneratingFunctionForAFive}) with the estimated values of the parameters \smash{$a_{0}$}, \smash{$a_{1}$}, \smash{$b_{1}$}. We have also plotted the standard Mar\v{c}enko-Pastur (Bai-Silverstein) \cite{SilversteinBai1995,MarcenkoPastur1967}
\subsubsection{Discussion}
The result is important for forecast design,
but more importantly, it provides information about the way macroeconomic variable
interact. The empirical data as compared to the spectral density equation suggest that a lot of eigenvalues, similarly to stock market data express marginal predictive content. One can suppose, that each of the economic time series contains important information about the collective movements, that cannot be gleaned from other time series. Alternatively, if we suppose that macro variables interact in the simplest low-dimensional way suggested by $\VARMAOneOne$ model,the conformity is nearly ideal (modulo the finite--size effects at the right edge of the spectrum). The economic time series express common response to the ''global shock'' process \ie each eigenvalue now  contains useful information about the values of the factors, that affect co--movement and hence useful information about the future behavior of economy. Thus, while many more eigenvalues appear to be
useful, the predictive component is apparently common to many series in a way suggested by our simplified $\VARMAOneOne$ model.
The above analysis on a real - complex systems example - i.e. Economy of Poland, for which we have assumed, that each of the time series under study is generated by the same type of univariate $\VARMA$ process reveals a stunning fact,  that again the flawless correspondence between theoretical  spectral density and empirical data is found. We are in the position, where we cannot reject the hypothesis, there are indeed no autocorrelations among macroeconomic time series. One may also argue, that all these time series are closely bounded  with the process, which we will identify as ''global shock-process''\ie all time series represent the global response complex system under study, to a distortion and its adjustment to equilibrium state. This process is of univariate $\VARMA$ \ie $ARMA(1,1)$ type with hidden structure, that has to be revealed based on historical time series data. This crude empirical study allows potentially for variety of extensions. At least two are possible.  First, the approach used here identifies the underlying factors only up to a linear transformation, making economic interpretation of the factors
themselves difficult. It would be interesting to be able to relate the factors more
directly to fundamental economic forces in the spirit of DSGE models. Secondly, our theoretical result covers only stationary models, but say nothing about integrated, cointegrated and co--trending variables. We know that common long-run
factors are important for describing macroeconomic data, and theory needs to be
developed to handle these features in a large model framework.


\section{Unraveling Internal Temporal Correlations via SVD Technique}
In order to investigate the temporal properties of the internal correlations between two data sets one is often interested not only in the analysis of it's static properties given by Pearson estimator (\ref{eq:EstimatorcDefinition}), $\mathbf{C}_X=\frac{1}{T}\mathbf{\mathbf{XX^T}}$ but more likely how this features behave over a certain period of time. Again the primary way to describe cross--correlations in a Gaussian framework is through the two--point covariance function (\ref{eq:CovarianceFunctionDefinition}),
\begin{equation}
\mathbf{C}_{i a , j b} \equiv \la X_{i a} X_{j b} \ra .
\end{equation}
Where $X_{i a} \equiv x_{i a} - \la x_{i a} \ra$ are mean adjusted data, that can be further collected into a rectangular $N \times T$ matrix $\mathbf{R}$. The average $\langle \ldots \rangle$ is understood as taken according to some probability distribution whose functional shape is stable over time, but whose parameters may be time--dependent.
In previous sections we have used a very simplified form of the two--point covariance function (\ref{eq:CovarianceFunctionDefinition}), namely with cross--covariances and auto--covariances factorized and non--random (\ref{eq:CovarianceFunctionFactorization}) ,
\begin{equation}
\mathbf{C}_{i a , j b} = C_{i j} A_{a b}
\end{equation}
(we have assembled coefficients into an $N \times N$ cross--covariance matrix $\mathbf{C}$ and a $T \times T$ auto--covariance matrix $\mathbf{A}$; both are taken symmetric and positive--definite). The matrix of ``temporal structure'' $\mathbf{A}$ is a way to model two temporal effects: the (weak, short--memory) lagged correlations between the returns , as well as the (stronger, long--memory) lagged correlations between the volatilities (weighting schemes, eg.EWMA \cite{PafkaPottersKondor2004}). On the other hand, the matrix of spatial correlations $\mathbf{C}$ models the hidden factors affecting the variables, thereby reflecting the structure of mutual dependencies of the complex system. The salient feature assumed so far, these two matrices were decoupled and the assumption about the Gaussianity of random variables provides crude approximation, that variances of all random variables always exist. This was sufficient to fully characterize the dependencies of the $X_{i a}$'s. However, in more realistic circumstances (\ie building efficient multivariate models,which help understanding the relation between a large number of possible causes and resulting effects) one is more interested in the situations, where the spatio--temporal structure does not factorize. Cross-correlations technique (sometimes alluded as ''time--lagged correlations technique''\index{time--lagged correlations estimator}) is most likely meets these critical requirements.
\begin{equation}\label{Timelaggedcorrelationsestimator}
\mathcal{C}_{i a , j a + \Delta}(\Delta) = \frac{1}{T}\sum_{a=1}^T X_{ia}X_{j a+\Delta}
\end{equation}
The precise answer boils down to how to separate the spectrum of such a covariance matrix in the large $N$, large $T$ limit (\ie thermodynamical limit), when one can make use of the power of FRV calculus (see \cite{ThurnerBiely2007} for a solution based on the circular symmetry of the problem and Gaussian approximation). In this chapter we will very closely follow the method presented in \cite{BouchaudLalouxAugustaMiceliPotters2007}, where the authors suggested to compare the singular value spectrum of the empirical rectangular $M\times N$ correlation matrix with a benchmark obtained using Random Matrix Theory results (c.f. \cite{EdelmanRao2005}), assuming there are no correlation between the variables.  For $T \to \infty$ at $N,M$ fixed, all singular values should
be zero, but this will not be true if $T$ is finite.
The singular value spectrum of this benchmark problem can in fact
be computed exactly in the limit where $N,M,T \to \infty$,
when the ratios $m=M/T$ and $n=N/T$ fixed. Since the original description is veiled, for pedagogical purposes we rederive all these results in the language of FRV presented in the table ~\ref{tab:FRVanalog}. Furthermore we extend the results obtained in \cite{Snarska2008} to meet encounter time-lagged correlations.
\subsection{Mathematical Formulation of a Problem}
Due to the works \cite{Granger2001,BollerslevEngleWooldridge1998,Sims1980} it is believed that, the system itself should determine the number of relevant input and output factors. In the simplest approach one would take all the possible input and output factors and systematically correlate them, hoping to unravel the hidden structure. This procedure swiftly  blow up with just few variables. The cross - equation correlation matrix contains all the information about contemporaneous correlation in a Vector model and may be its greatest strength and its greatest asset. Since no questionable a priori assumptions are imposed, fitting a Vector model allows data--set to speak for itself \ie find the relevant number of factors. Still without imposing any restrictions on the structure of the correlation matrix one cannot make a causal interpretation of the results. The theoretical study of high dimensional factor models is indeed actively pursued in literature \cite{Geweke1997,StockWatson2002-1,StockWatson2002-2,StockWatson2005,ForniHallinLippiRechlin2000,ForniHallinLippiRechlin2004,Bai2003,BaiNg2002}. The main aim of this chapter is to  present a method, which helps extract  highly non-trivial spatio--temporal correlations between two samples of non-equal size (i.e. input and output variables of large dimensionality), for these can be then treated as "natural" restrictions for the correlations matrix structure.
\subsubsection{Basic framework and notation}
We will divide all variables into two subsets \ie focus on $N$ input factors $X_a$ $(a=1,\dots, N)$ and $M$ output factors $Y_{\alpha}$ $(\alpha =1,\ldots,M)$ with the total number of observations being $T$. All time series are standardized to have zero mean and unit variance. The data can be completely different or be the same variables but observed at different times.  First one has to remove potential correlations inside each subset, otherwise it may interfere with the out-of-sample signal. To remove the correlations inside each sample we form two correlation matrices,which contain information about in-the-sample correlations.
\begin{equation}\label{internal}
\mathbf{C_X}=\frac{1}{T}XX^T \qquad \mathbf{C_Y}=\frac{1}{T}YY^T
\end{equation}
The matrices are then diagonalized,provided $T > N,M$, and the empirical spectrum is compared to the theoretical Mar\v{c}enko-Pastur spectrum \cite{MarcenkoPastur1967,LalouxCizeauBouchaudPotters1999,BurdaJurkiewicz2004,BurdaGorlichJaroszJurkiewicz2004}in order to unravel statistically significant factors.The eigenvalues,which lie much below the lower edge of the Mar\v{c}enko-Pastur spectrum represent the redundant factors, rejected by the system, so one can exclude them from further study and in this manner reduce somewhat the dimensionality of the problem, by removing possibly spurious correlations. Having found all eigenvectors and eigenvalues, one can then construct a set of uncorrelated unit variance input variables $\hat{X}$ and output variables  $\hat{Y}$.
\begin{equation}\label{result}
\hat{X}_{at}=\frac{1}{\sqrt{T\lambda_a}}V^TX_t\qquad \hat{Y}_{\alpha t}=\frac{1}{\sqrt{T\lambda_{\alpha}}}U^TY_t
\end{equation}
where $V,U$,  $\lambda_a$, $\lambda_{\alpha}$ are the corresponding eigenvectors and eigenvalues of $C_X$ , $C_Y$ respectively. It is obvious, that $C_{\hat X}=\hat X \hat X^T$ and $C_{\hat Y}=\hat Y \hat Y^T$ are  identity matrices, of dimension, respectively, $N$ and $M$. Using general property of diagonalization, this means that the $T \times T$ matrices $D_{\hat X}=\hat X^T \hat X$ and $D_{\hat Y}=\hat Y^T \hat Y$
have exactly $N$ (resp. $M$) eigenvalues equal to $1$ and $T-N$ (resp. $T-M$) equal to zero. These non-zero eigenvalues are randomly arranged on a diagonal.
Finally we can reproduce the asymmetric $M\times N$ cross-correlation matrix $G$ between the $\hat{Y}$ and $\hat{X}$:
\begin{equation}
G=\hat{Y}\hat{X}^T
\end{equation}
which includes only the correlations between input and output factors. In general the spectrum of such a matrix is complex, but we will use the singular value decomposition (SVD) technique (c.f. \cite{FriedbergInselSpence2002})  to find the empirical spectrum of eigenvalues.
\subsubsection{The Singular Value Decomposition}
The singular value spectrum represent the strength of cross-correlations between input and output factors. Suppose $G$ is an $M\times N$ matrix whose entries are either real or complex numbers. Then there exists a factorization of the form
\begin{equation}\label{SVDdecomposition}
    G = U\Sigma V^{\dagger}
\end{equation}
where $U$ is an $M\times M$ unitary matrix. The columns of U form a set of orthonormal ''output'' basis vector directions for $G$ - these are the eigenvectors of $G^{\dagger}G$. $\Sigma$ is $M\times N$ diagonal matrix with nonnegative real numbers on the diagonal,which can be thought of as scalar ''gain controls'' by which each corresponding input is multiplied to give a corresponding output. These are the square roots of the eigenvalues of $GG^{\dagger}$ and $G^{\dagger}G$ that correspond with the same columns in U and V.
 and $V^{\dagger}$ denotes the conjugate transpose of $V$, an $N\times N$ unitary matrix,whose columns form a set of orthonormal ''input'' or vector directions for $G$. These are the eigenvectors of $GG^{\dagger}$.
A common convention for the SVD decomposition is to order the diagonal entries $\Sigma_{i,i}$ in descending order. In this case, the diagonal matrix $\Sigma$ is uniquely determined by $G$ (though the matrices $U$ and $V$ are not). The diagonal entries of $\Sigma$ are known as the singular values of $G$.
\subsection{Singular values from free random matrix theory}
In order to evaluate these singular eigenvalues, assume without loss of generality $M < N$. The trick is to consider the matrix $M \times M$ matrix $GG^T$ (or the $N \times N$ matrix $G^TG$ if $M > N$), which is symmetrical and has $M$ positive eigenvalues, each of which being equal to the square of a singular value of $G$ itself. Furthermore use the cyclic properties of the trace. Then non-zero eigenvalues of \[GG^T=\hat Y \hat X^T\hat X \hat Y^T\] are  then the same (up to the zero modes) as those of the $T \times T$ matrix \[\mathbf{D} = D_{\hat X} D_{\hat Y}=\hat X^T \hat X \hat Y^T\hat Y\] obtained by swapping the position of $\hat Y$ from first to last.
In the limit $N,M,T\rightarrow\infty$ where the $\hat X$'s and the $\hat Y$'s are independent from each other, the two matrices
$D_{\hat X}$ and $D_{\hat Y}$ are mutually free \cite{Voiculescu1991}, and we can use the results from FRV, where given the spectral density of each individual matrix, one is able to construct the spectrum of the product or sum of them.
\subsubsection{FRV Algorithm for Cross-correlation matrix}
As usual we will start with constructing the Green's function for matrices $\mathbf{D}_{\hat{X}}$ and $\mathbf{D}_{\hat{Y}}$. Each of these matrices, have off-diagonal elements equal to zero, while on diagonal a set of $M$ (or $N$ respectively) randomly distributed eigenvalues equal to $1$
\begin{equation}
\begin{array}{rcl}
G_{D_{\hat{X}}}&=&\frac{1}{T}\left(\frac{M}{z-1} +\frac{T-M}{z}\right)=\frac{m}{z-1} +\frac{1-m}{z}\quad m=\frac{M}{T}; \\ G_{D_{\hat{Y}}}&=&\frac{1}{T}\left(\frac{N}{z-1} +\frac{T-N}{z}\right)=\frac{n}{z-1} +\frac{1-n}{z}\quad n=\frac{N}{T};
\end{array}
\end{equation}
Then:
\begin{equation}
\mathbf{S}_{D_{\hat{X}}\cdot D_{\hat{Y}}}(z)=\mathbf{S}_{D_{\hat{X}}}(z)\cdot \mathbf{S}_{D_{\hat{Y}}}(z)
\end{equation}
or equivalently
\begin{equation}
\frac{1+z}{z}\mathbf{N}_{D_{\hat{X}}\cdot D_{\hat{Y}}}(z)=\mathbf{N}_{D_{\hat{X}}}(z)\cdot \mathbf{N}_{D_{\hat{Y}}}(z),
\end{equation}
where:
\[\mathbf{S}_{x}(z)=\frac{1+z}{z}\mathbf{\chi}_{x}(z) \qquad \mathbf{N}_{x}(z)=\frac{1}{\mathbf{\chi}_{x}(z)}\qquad \mathbf{N_{x}(z)}G\left(\mathbf{N_{x}(z)}\right)-1=z
\]
From this, one easily obtains:
\begin{equation}\begin{array}{rcl}
\mathbf{N}_{D_{\hat{X}}(z)}\left(\frac{m}{\mathbf{N}_{D_{\hat{X}}(z)}-1} +\frac{1-m}{\mathbf{N}_{D_{\hat{X}}}(z)}\right)-1&=&z\\
\frac{\mathbf{N}_{D_{\hat{X}}}(z)m}{\mathbf{N}_{D_{\hat{X}}}(z)-1} +1 -m -1 &=&z
\end{array}
\end{equation}
\begin{equation}
\mathbf{N}_{D_{\hat{X}}}(z)=\frac{m+z}{z} \qquad \mathbf{N}_{D_{\hat{Y}}}(z)=\frac{n+z}{z}
\end{equation}
\begin{equation}\label{fullNtransform}
\mathbf{N}_{D_{\hat{X}}}\mathbf{N}_{\cdot D_{\hat{Y}}}(z) =\frac{(m+z)(n+z)}{z^2}
\end{equation}
and one readily gets the $N$--transform for the matrix $\mathbf{D_{\hat{X}}\cdot D_{\hat{Y}}}$
\begin{equation}\label{fullStransform}
\mathbf{N}_{D_{\hat{X}}\cdot D_{\hat{Y}}}(z)=\frac{(m+z)(n+z)}{z(1+z)}
\end{equation}
Inverting functionally (\ref{fullStransform})
\begin{equation}
\mathbf{N}_{D_{\hat{X}}\cdot D_{\hat{Y}}}(z)\mathbf{G}_{\mathbf{D}}\left(N_{D_{\hat{X}}\cdot D_{\hat{Y}}}(z)\right)=z+1
\end{equation}
\ie solving the second order equation in $z$, one is able to find the Green's function of a product $D_{\hat{X}}\cdot D_{\hat{Y}}$
\begin{equation}
\begin{array}{rcl}
0&=&z^2(1-N(z)) +(n+m-N(z))z+mn \\
G(N(z))&=&\frac{2-(n+m+N(z))-\sqrt{(n+m-N(z))^2-4(1-N(z))mn}}{2N(z)(1-N(z))},
\end{array}
\end{equation}
where we have omitted the subscripts for brevity.
Subsequently mean spectral density is obtained from the standard relation
\begin{equation}
\lim_{\epsilon\to 0^+}\frac{1}{\lambda+i\epsilon}=PV\left(\frac{1}{\lambda}\right) -i\pi\delta(\lambda) \Rightarrow\rho_{\mathbf{D}}(\lambda)=-\frac{1}{\pi}\im G_{\mathbf{D}}(\lambda +i\epsilon).
\end{equation}
The final result \ie the benchmark case where all (standardized)
variables $X$ and $Y$ are uncorrelated, meaning that the ensemble average  $E(C_X)=E(XX^T)$ and $E(C_Y)=E(YY^T)$
are equal to the unit matrix, whereas the ensemble average cross-correlation $E(G)=E(YX^T)$ is
identically zero, reads as in original paper \cite{BouchaudLalouxAugustaMiceliPotters2007}:
$$\rho_{\mathbf{D}}(\lambda) = \max(1-n,1-m) \delta(\lambda) + \max(m+n-1,0) \delta(\lambda-1) +$$
\begin{equation}
+\frac{\re \sqrt{(\lambda^2-s_-)(\lambda_+-s^2)}}{\pi \lambda(1-\lambda^2)}
\end{equation}
where $s_{\pm} = n+m-2mn \pm 2 \sqrt{mn(1-n)(1-m)}$ are the two positive roots of the quadratic expression
under the square root
\begin{figure}[!h]
\begin{center}
\includegraphics[width=0.7\textwidth]{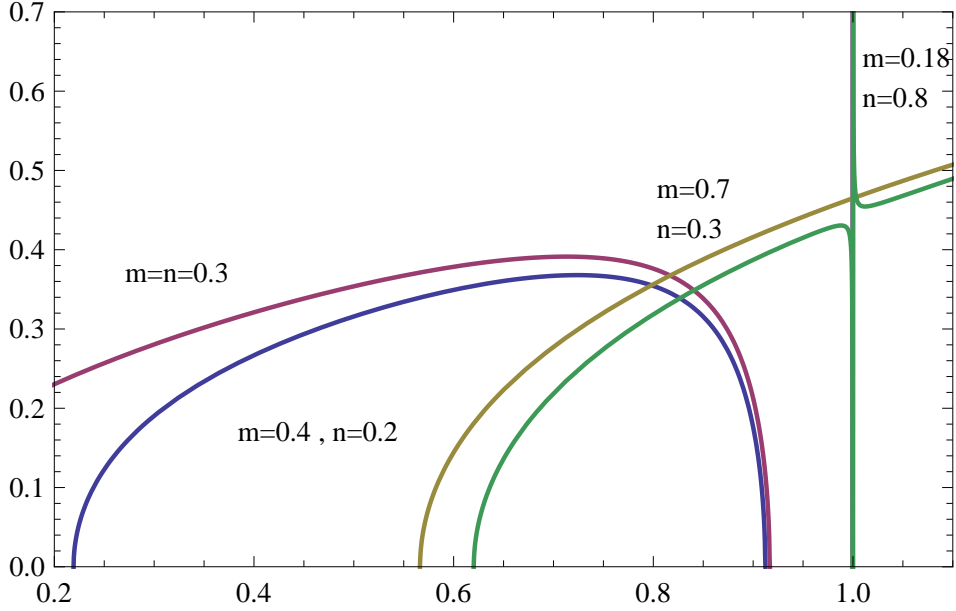}
\end{center}
\caption{
Simulation of a continuous part of the theoretical random singular value
spectrum $\rho(\lambda)$ for different values of $n$ and $m$.It is obvious to see
that $\lambda_+ \leq 1$ for all values of $n,m <1$. The upper
bound is reached only when $n+m=1$, in which case the upper edge of the
singular value band touches $\lambda=1$ \ie for
$n=m$ the spectrum extends down to $\lambda=0$, whereas for $n+m \to 1$,
the spectrum develops a $(1-\lambda)^{-1/2}$ singularity, just before the
appearance of a $\delta$ peak at $\lambda=1$ of weight $n+m-1$.
}
\label{SVDtheoreticalspectrum}
\end{figure}
It is easy to  discover the fact, that in the limit $T \to \infty$ at fixed $N$, $M$,
all singular values collapse to zero, as they should
since there is no true correlations between $X$ and $Y$;
the allowed band in the limit $n,m \to 0$ becomes:
\begin{equation}\lambda \in \left[|\sqrt{m}-\sqrt{n}|,\sqrt{m}+\sqrt{n}\right].
\end{equation}
When $n \to m$, the support becomes $\lambda \in [0, 2\sqrt{m(1-m)}]$
(plus a $\delta$ function at $\lambda=1$ when $n+m > 1$),  while when $m=1$,
the whole band collapses to a $\delta$ function at
$\lambda=\sqrt{1-n}$. For $n+m \to 1^-$ there is an initial singularity of $\rho(\lambda)$
$\lambda =1$ diverging as $(1-\lambda)^{-1/2}$. Ultimately $m \to 0$ at fixed $n$,
one finds that the whole band collapses again to a $\delta$ function at
$\lambda = \sqrt{n}$.
\subsubsection{SVD cleaning technique and the $MP^2$ case}
The results from the previous section were obtained under belief there were no correlations between input and output samples of infinite sizes. However, for a given finite size sample, the eigenvalues of $C_X$ and $C_Y$
will differ from unit, and the singular values of $G$ will not be zero and instead {\it cross}-correlations between
input and output variables are involved.
The SVD spectrum in that case is the convolution of two Mar\v{c}enko-Pastur \cite{MarcenkoPastur1967} distributions with parameters $m$ and $n$, respectively, which reads, for $r=n,m < 1$:
\begin{equation}
\rho_{MP}(\lambda)= \frac{1}{2 \pi \beta \lambda} \re \sqrt{(\lambda-\lambda_-)(\lambda_+
-\lambda)}
\end{equation}
with $\lambda_\pm=(1 \pm \sqrt{r})^2$
The $N$-transform of this density takes a particularly simple form (cf. \cite{BurdaJaroszJurkiewiczNowakPappZahed2009} for an exact derivation)
\begin{equation}
\mathbf{N}_{MP}(z)=\frac{1+z}{1+rz}
\end{equation}
The singular values of $G$ are obtained as the square-root of
the eigenvalues of $D=X^TXY^TY$. Under assumption, that $X^TX$ and $Y^TY$ are mutually free,  after having noted that the $N$-transform of the $T\times T$ matrices $X^TX$ and $Y^TY$ are now given by:
\begin{equation}
\mathbf{N}(z)=\frac{(1+z)(1+rz)}{rz}
\end{equation}
one can again use the multiplication rule of $N$-transforms and finds the Green's function of $D$ by solving the following
cubic equation for $z$:
\begin{equation}\label{cubicMP2}
(1+z)(1+nz)(1+mz)\mathbf{N}(z)-mnz=0
\end{equation}
which with little effort can be solved analytically. Then Green's function is readily obtained by inserting the solution of the eq.(\ref{cubicMP2})
\begin{equation}
G(\BN(z))=\frac{z(\BN(z)) +1}{\BN(z)} \Rightarrow \rho(\lambda^2)=-\frac{1}{\pi}\im G(\lambda +i\epsilon)
\end{equation}
This will lead to a rather complicated form of the final function
\begin{equation}
\rho(\lambda)=\left(\frac{2}{\theta(\lambda^2)}\right)^{1/3}\frac{3^{-1/2}}{\pi\lambda}\left(2^{-2/3} +\varphi(\lambda^2)\right)
\end{equation}
where
\[\varphi(\lambda^2)=2-3m(1-m)-3n(1-n)-3mn(n+m-4)+2(m^3+n^3)+9\lambda^2(1+m+n)\]
\[\theta(\lambda^2)=\varphi(\lambda^2) -\sqrt{\varphi(\lambda^2) -4(1+m^2+n^2-mn-m-n+3\lambda^2)^3}\]
\subsection{Example from the Data}
The last decade has been a witness of an enormous progress in the development of small-scale macroeconomic models.  It's not too much an overstatement to say, that the statistical analysis of VAR models, Kalman filter models etc. is nowadays complete. The major issue with these models is that they can accurately approximate small number of time series only. On the other hand Central Banks must construct their forecasts in rich data environment \cite{BernankeBoivin2003}. This mismatch between standard macroeconometric models and real world
practice has led to unfortunate consequences. Forecasters have had to rely on
informal methods to distill information from the available data, and their published forecasts reflect considerable judgement in place of formal statistical analysis. Forecasts
are impossible to reproduce, and this makes economic forecasting a largely non-scientific
activity \ie formal small-scale models have little effect on day-to-day policy
decisions, making these decisions more ad hoc and less predicable than if guided by the
kind of empirical analysis that follows from careful statistical modeling.
The goal of this research is to use the wide range of economic
variables that practical forecasters and macroeconomic policymakers have found useful,
and establish a direction that explicitly
incorporates information from a large number of macroeconomic variables into formal
statistical models.
We have focused on two different data sets, namely Polish macroeconomic data and generated set of data, where temporal cross - correlations are introduced by definition.The full data set is the same as it was used in previous chapter.
\subsubsection{Polish Macroeconomic data revisited}
Poland is and interesting emerging market with unique social and business activity in the process of rapid growth and industrialization. We hope our analysis might be helpful in understanding the factors that helped Poland to survive during the 2008 crisis. The main problem to be solved is to choose the correct variables to include.
This is the familiar problem of variable selection in regression analysis. Economic theory
is of some help, but usually suggests large categories of variables (money, interest rates,
wages, stock prices, etc.) and the choice of a specific subset of variables then becomes an open problem.
The analysis began with checking, whether the method described in \cite{BouchaudLalouxAugustaMiceliPotters2007} is relevant for describing the relation between the inflation indexes for Polish macroeconomic indexes and other Polish macroeconomic data published by different government and non-government agencies. A consumer price index (CPI) is a measure estimating the average price of consumer goods and services purchased by households. A consumer price index measures a price change for a constant market basket of goods and services from one period to the next within the same area (city, region, or nation). It is a price index determined by measuring the price of a standard group of goods meant to represent the typical market basket of a typical urban consumer. The percent change in the CPI is a measure estimating inflation. It is commonly viewed as the indicator not only the measure of inflation, but rather the indicates the change of costs of 	maintenance. The data set represent a wide range of macroeconomic activity and were initially transformed to ensure stationarity and diminish the effects of seasonal components.
The same data set we have already analyzed in the first part of the paper  and the detailed list of all time series can be obtained from the author upon request.  This time, the whole set of  $52$ time series,observed on a monthly basis between $Jan-2000$ and $Oct-2009$  ($T=118$) was divided into two subsets \ie
 \begin{itemize}
\item We have used monthly $M=15$ changes of different CPI indicators as our predicted variables (i.e. output sample $Y$)
 \item The input sample $X$ consisted of $N=37$ monthly changes of economic indicators (eg. sectoral employment, foreign exchange reserves, PPI's) as explanatory variables.
 \end{itemize}
 The data were standardized and mean adjusted, but following the general idea of \cite{BouchaudLalouxAugustaMiceliPotters2007} the input and output samples' factors were not selected very carefully, so the data could speak for themselves and system could be able to select the optimal combination of variables.
\begin{figure}[!h]\begin{center}
  \includegraphics[width=0.45\textwidth]{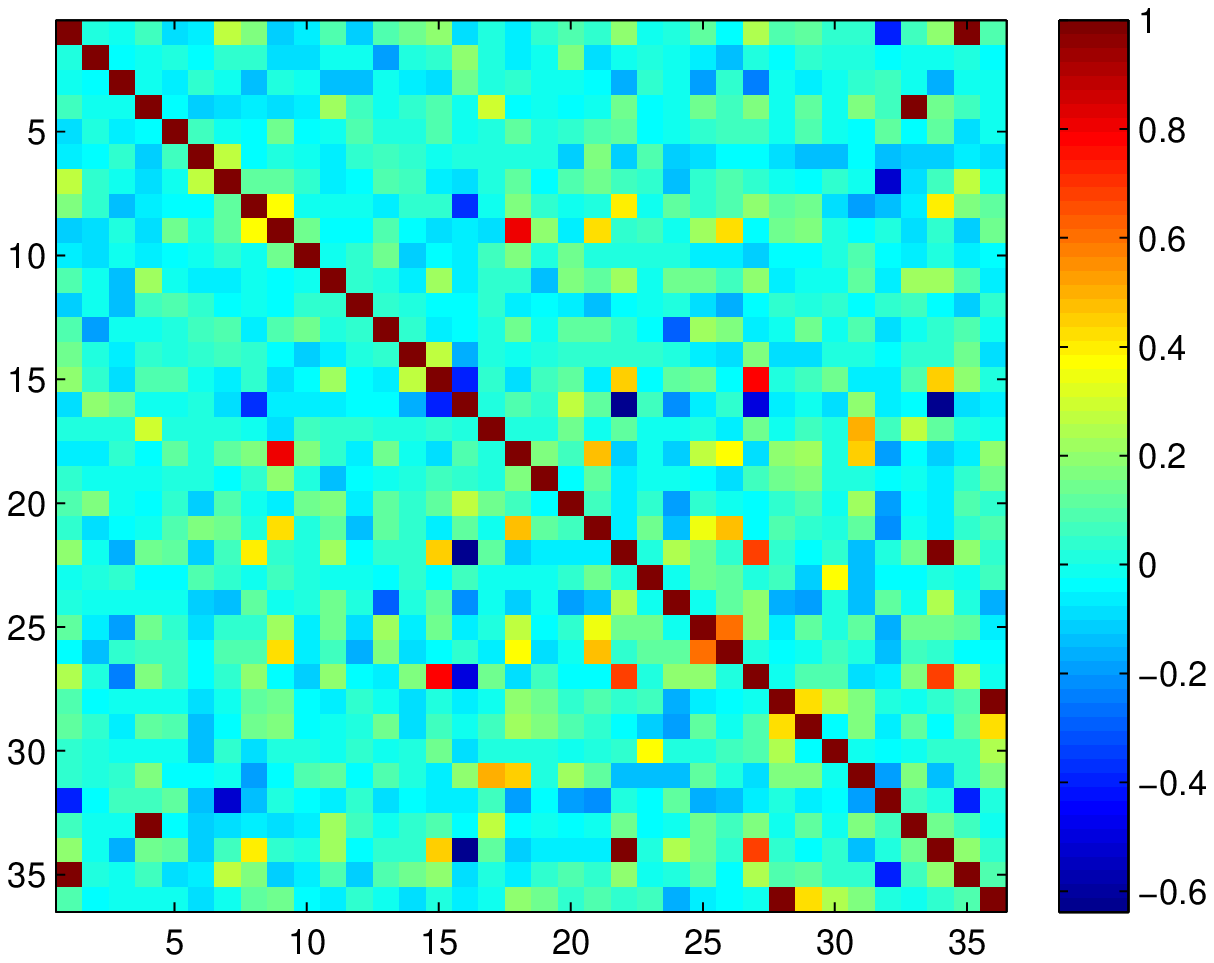}
  \includegraphics [width=0.45\textwidth]{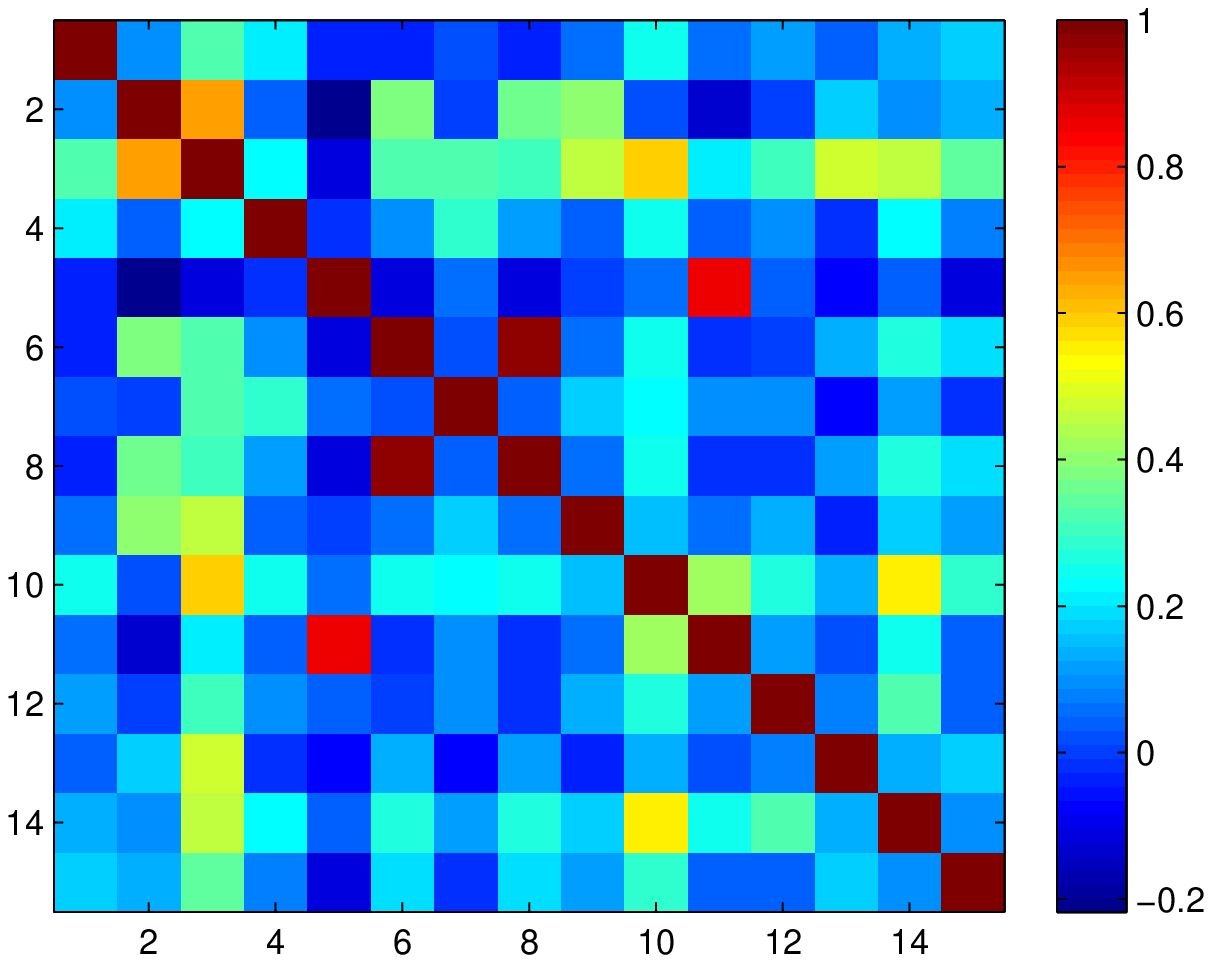}
  \caption[Pearson correlation matrices representing generic in-the-sample correlations]{Correlation matrices representing generic in-the-sample correlations.The data were mean adjusted and standardized. In a perfect situation, one is expecting that cross--correlations tends to zero, however still nontrivial correlations are present. LEFT:Matrix with $37$ input variables $X$.RIGHT: Matrix with $15$ input variables $Y$ - components of CPI.}\label{fig:XYspectra}
  \end{center}
\end{figure}
 The resulting diagrams (see Fig.\ref{fig:XYspectra}) now demonstrate, that even standardized and converted to stationary time series may represent nontrivial in-the-sample correlations. Short -- term economic forecasts build from these type data in consequence may be poor and show no sign of improving over time.
The next step involved cleaning internal correlations in each sample. To do it, we have used equation (\ref{internal}). The effective matrices were then diagonalized and two sets of internally uncorrelated data were prepared.
\subsubsection{Results for Equal-time spectra}
\begin{figure}[!h]
 \begin{center}
  \includegraphics[width=0.45\textwidth]{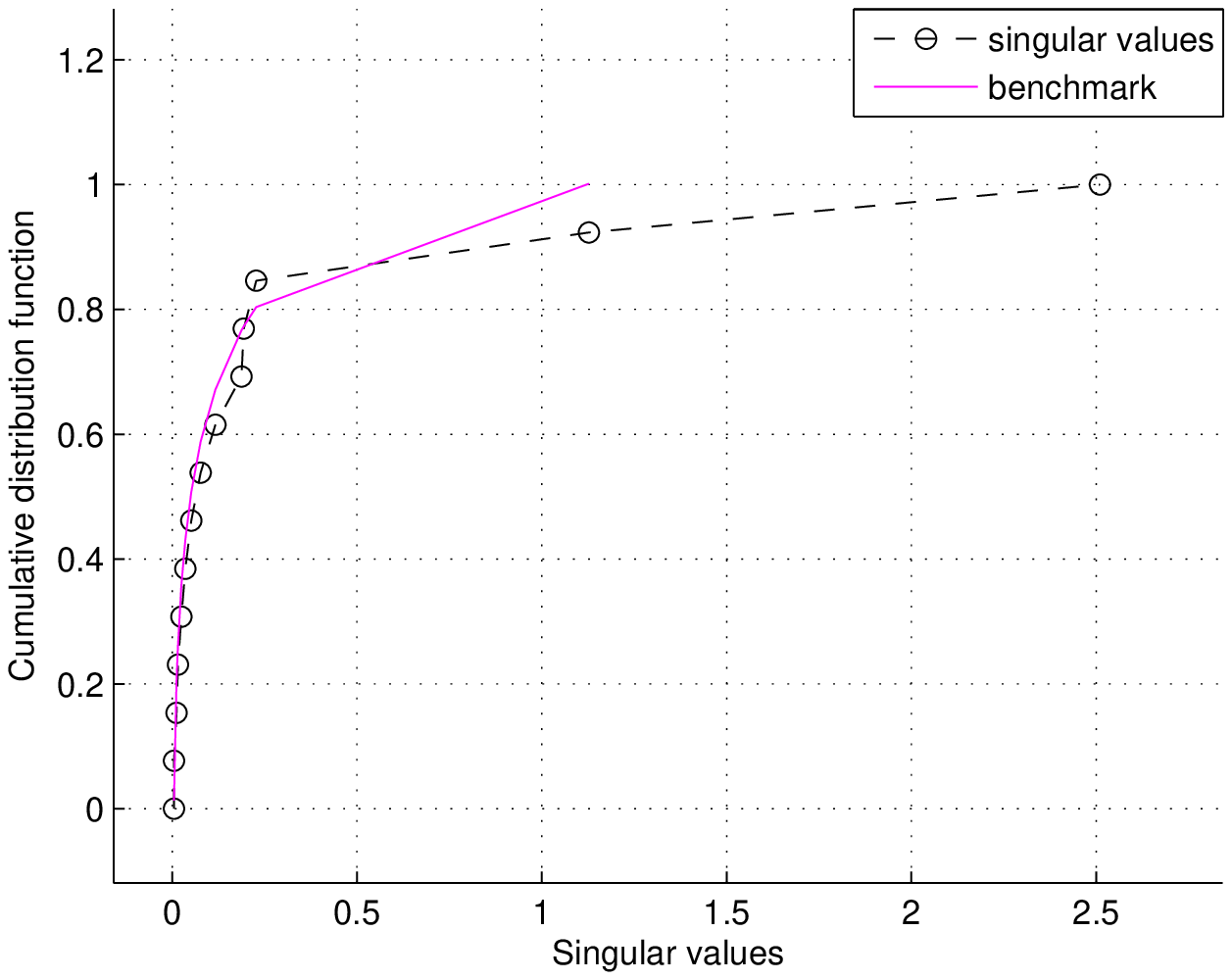}\includegraphics [width=0.45\textwidth]{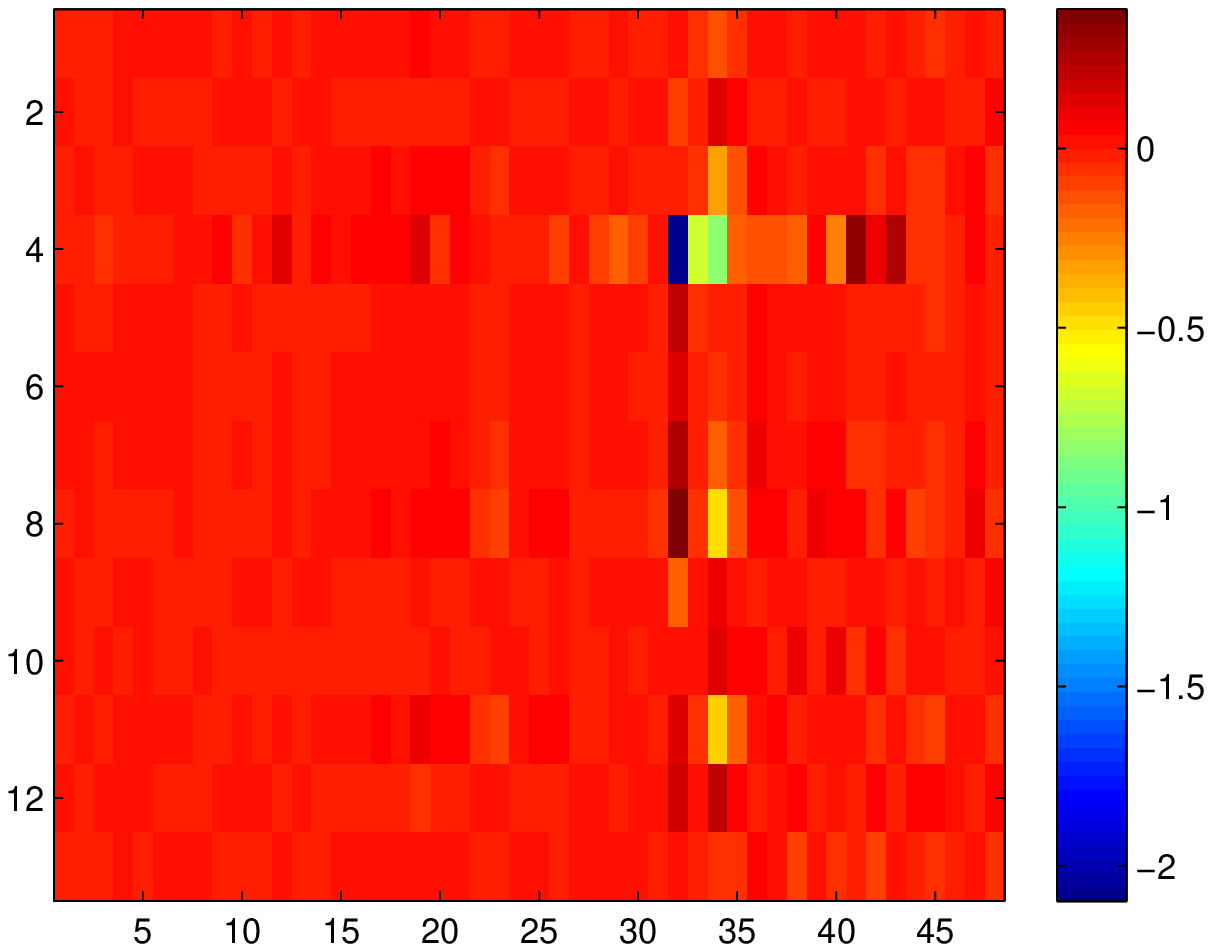}\\
  \includegraphics [width=0.45\textwidth]{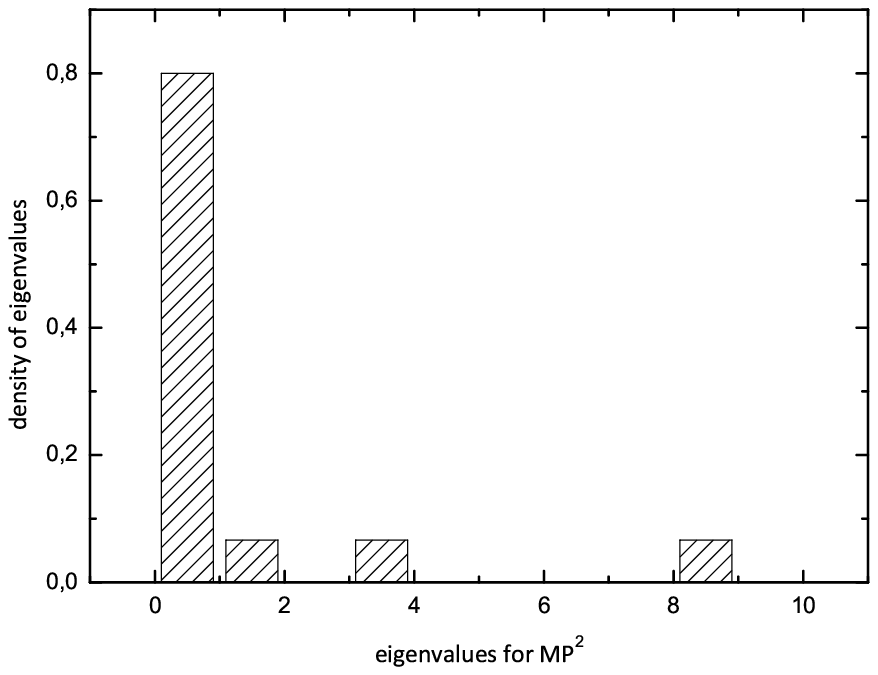}\includegraphics[width=0.45\textwidth]{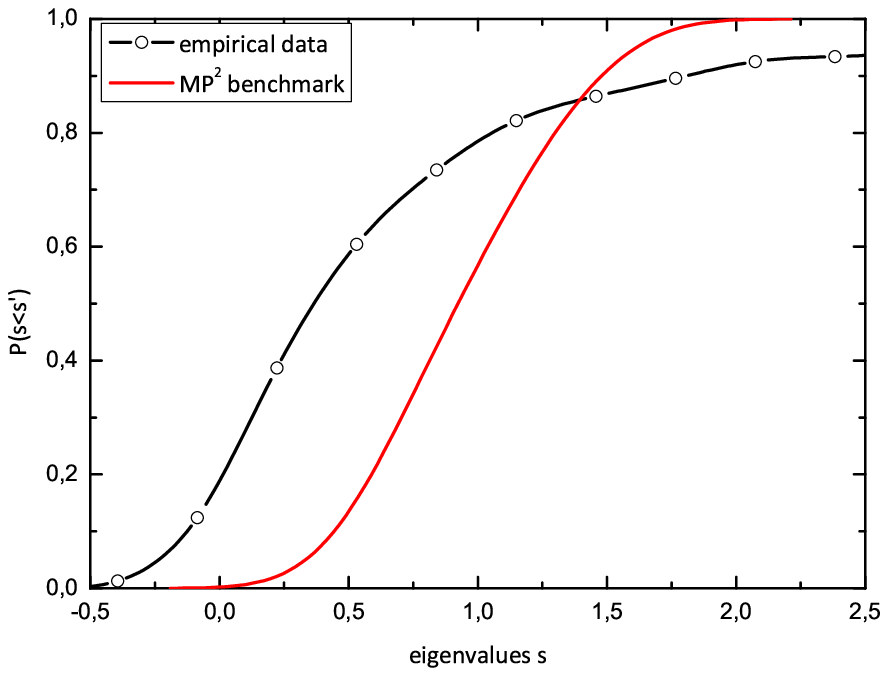}
\caption[Comparison of Cleaning techniques for equal--time SVD spectra]{Comparison of cleaning techniques for equal--time SVD spectra. ABOVE: Cumulative singular density and ''heat map'' for the cleaned problem. BELOW LEFT: Empirical density of eigenvalues in the $MP^2$ framework.BELOW RIGHT: Benchmark calculated according to $MP^2$ spectrum.}\label{fig:initialanalysis}
\end{center}
\end{figure}

From the uncorrelated data we create the rectangular matrix $G$ and diagonalize it to calculate singular eigenvalues. Finally we have used the benchmark calculated in equation (\ref{result}) to compare the data with the predicted eigenvalue density. For the same data sets we have also created the set of correlated samples \ie the set, where internal spatial cross--correlations were not a-priori removed (see Fig. \ref{fig:initialanalysis}). Apparently there is enough idiosyncratic variation in standard activity measures like the
unemployment rate and capacity utilization, that removing noisy components from these might provide a clearer picture of factors affecting inflation. We have excluded from further analysis series responsible for reference NBP bill rate balance of payments, and from the set of explanatory variables ordinary measures of inflation - CPI in food sector, beverages and tobacco ans services.
This approach allows us to directly reproduce temporal cross--correlations.\\ The lack of symmetry condition endure us to focus only on out-of-the-sample correlations without mixing them with inner ones and to study temporal properties of such matrix. The results show, that there exists some singular eigenvalues, which do not fit the benchmark. Among them, the highest singular eigenvalue $s_1=2.5$ and the corresponding singular eigenvector, represent standard negative correlation between expenses for electricity and net balance of payments in the energy and positive correlation between CPI in health sector and unemployment. In our approach we can not only observe this relations, but also interpret them in terms of causality. That is, larger unemployment rate causes increase in the CPI. There are other non-trivial relations between eg. CPI in telecommunication sector and foreign exchange reserves. All of these correlations are however well known from textbooks or can be easily explained by means of classical economic theory. When some of the eigenvalues become strongly related, zero modes emerge - clearly the majority (around 80\% of all eigenvalues are concentrated close to zero, meaning that there are strong spatial correlations inside $X$ and $Y$ data set. If we are to use the $MP^2$ benchmark then it is clear, that empirical spectrum is affected by idiosyncratic components, again confirming, that spatial structure strongly interferes with temporal, and it is crucial to ''remove'' redundant factors to avoid spurious (confunding) correlations.
\subsubsection{Solution for lagged spectra}
A natural way to examine macroeconomic data is via factor models. In previous section we have assumed that the inflation can be accurately explained by the factor model using relatively small number of latent variables.
Pushing the factor model one step further, these latent factors might also explain the predictive relationship between current values of variables and those of the previous month.
The next step of our study involved shifting the input and output data set by one observation (one month). The $Y$ were calculate from $t=2,\ldots,118$ and $X$'s for $(t=1,\ldots,117)$. We were motivated by the common belief, that it is the ''yesterday'' shock, that affects ''today'' change \ie there is some persistency and long memory within this system. This is the same approach, that underlies the idea of $\VARMA$ processes \cite{Lutkepohl2005}.
 \begin{figure}[!h]
 \begin{center}
  \includegraphics[width=0.45\textwidth]{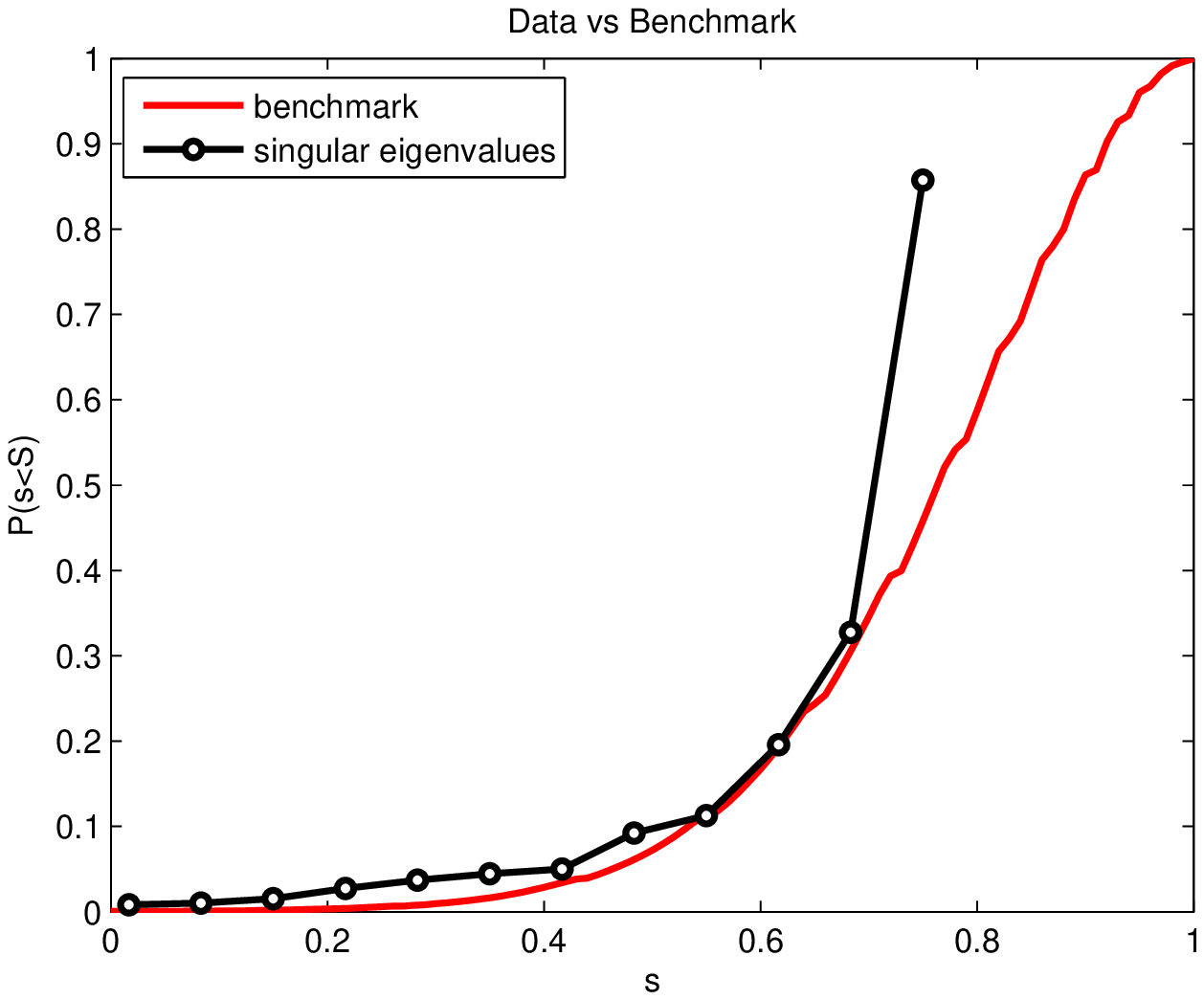}\includegraphics [width=0.45\textwidth]{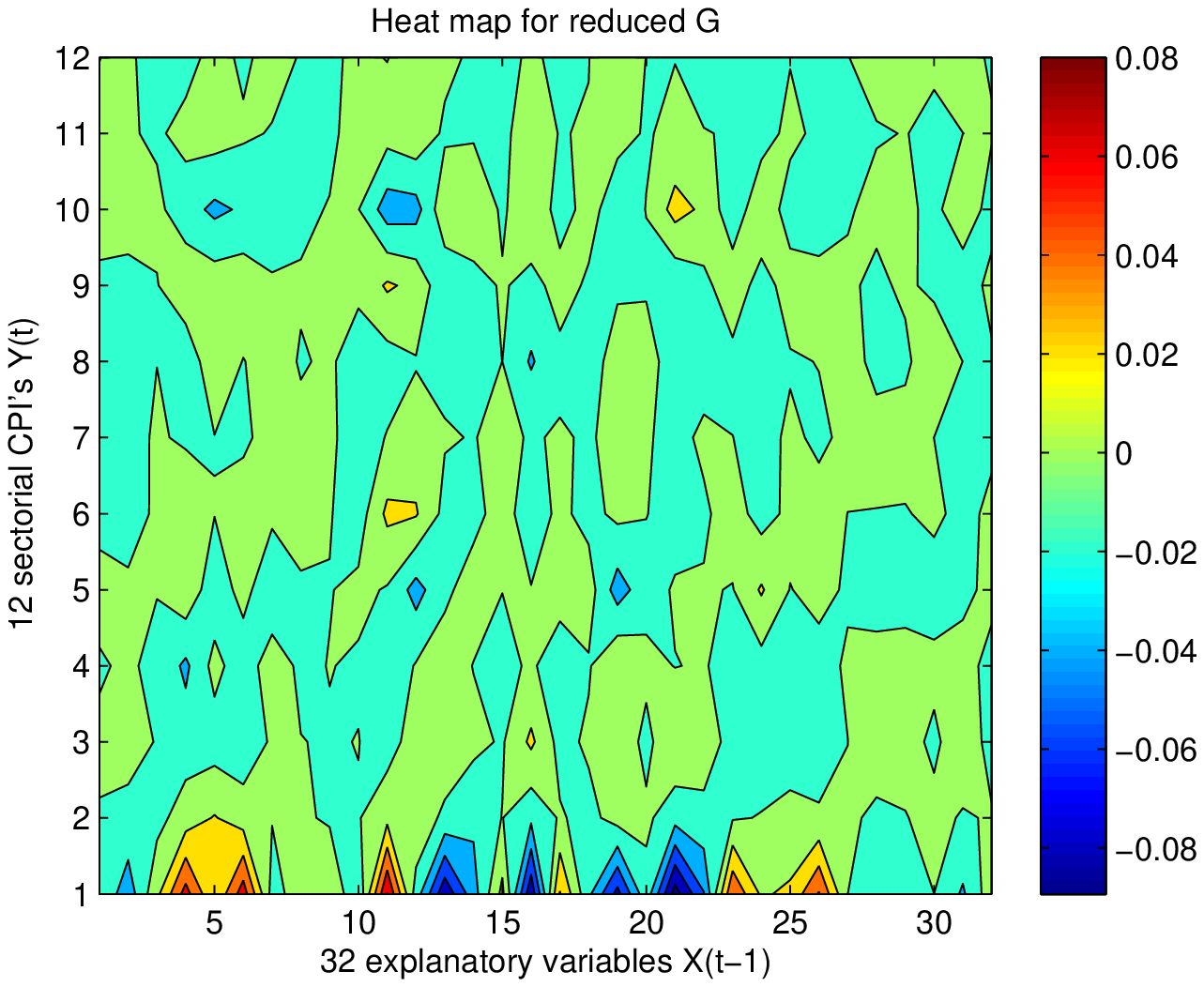}
\caption[Results for lagged spectra]{Results for lagged-by-one-month spectra.LEFT: There is a bunch of singular values that do not fit the benchmark. The largest $s\approx 0.77$ represents the same correlation structure as within unshifted framework.RIGHT: Note, that now we can see whole bunch of  islands of non-trivial  factors, that affect CPI's.}\label{fig:XYshifted}
\end{center}
\end{figure}
The temporal structure (Fig.\ref{fig:XYshifted}) manifests itself via the existence of significant non--symmetric relation (represented by one singular eigenvalue, that does not fit the benchmark) between data sets $X$ and $Y$, that are shifted by one month. It is easy to notice, that only few factors are responsible for the model's performance.
\begin{table}[!h]
\begin{center}
\begin{tabular}{|p{0.25\textwidth}|p{0.4\textwidth}|p{0.25\textwidth}|}\hline
$Y$&$X$&Type of correlation\tabularnewline\hline
CPI in communication sector&completed dwellings&negative\tabularnewline
&net balance of payments of goods&\tabularnewline
&$M3$ money aggregate&\tabularnewline
&Employment in manufacturing sector&\tabularnewline\cline{2-3}
&employment in enterprize sector&positive\tabularnewline
&Direct investments&\tabularnewline
&Foreign exchange reserves&\tabularnewline
&Official reserve assets&\tabularnewline
&New heavy trucks registration&\tabularnewline
&Balance of payments - services&\tabularnewline\hline
CPI in clothing sector&Total export&negative\tabularnewline\hline
CPI in restaurants and hotels sector&Foreign exchange reserves&positive\tabularnewline\hline
CPI in transport sector& Foreign exchange reserves&positive\tabularnewline\cline{2-3}
&Total production in manufacturing sector&negative\tabularnewline
&Total export&\tabularnewline \hline
\end{tabular}\end{center}
\caption{Factors affecting the set of output variables for lagged spectra.}\label{factors}
\end{table}
CPI in telecommunication sector is affected by the largest number of possible explanatory variables (c.f. Table \ref{factors}). Among them the most unexpected is the correlation with heavy trucks. Two or three
factors are useful for some categories of series, but only a single factor is responsible for
the predictability of prices in all sectors.
Apparently, the first factor is foreign exchange reserves level, and the
results say that it is an important predictor of future prices in telecommunication, manufacturing and transport sector.
We can say that when forecasting inflation a large model might be a clue, but if we remove redundant factors the inflation can be forecasted by using simple measures of real activity like the unemployment
rate, industrial production or capacity utilization.
While the first factor is easy to interpret, a complete understanding of the
results requires an understanding of other factors as well. Unfortunately, their interpretation and role in explaining future changes in the consumer prices is an open question.
\subsection{Conclusions}
 We will now recap this illustrative study with few comments:
 \begin{itemize}
 \item In general both input and output data sets may represent highly complex correlation structure strongly interfered by redundant noisy factors. This significant amount of noise need to be carefully eliminated by performing initial decoupling of spatial correlations, so these large matrices become mutually free.
 \item  This is again precisely the case when FRV approach ''takes the stage'' and reduces the solution to few lines.\
 \item  The procedure tested on real data within the case of unshifted variables hasn't show any significant improvement in comparison to standard factor analysis known in econometric literature for similar data sets\cite{StockWatson1999}. For data lagged by one observation we have however recovered the sea of different non--trivial relations, and it might be interesting to compare these results from a more general perspective of factor models, however no implicitly close approach was found in the literature.
 \end{itemize}

\begin{acknowledgments}
This work has been supported by the Polish Ministry of Science Grant No.~N~N202~229137 (2009--2012).
\end{acknowledgments}



\begin{thebibliography}{99}

\bibitem{Wold1938}
{\sc Wold, H.}
\newblock A study in the analysis of stationary time series.
\newblock Tech. rep., Almquist and Wiksell, Uppsala, 1938.
\newblock
  \href{http://www.dli.ernet.in/scripts/FullindexDefault.htm?path1=/data/upload/0012/812&first=1&last=246&barcode=99999990012807}{[DOI:99999990012807]}.

\bibitem{Sims1980} {\sc Sims C. A.}, \emph{Macroeconomics and reality}, Econometrica \textbf{48} (1980) 1.

\bibitem{Lutkepohl2005} L\"{u}tkepohl H., \emph{New Introduction to Multiple Time Series Analysis}, Springer Verlag, Berlin, 2005.

\bibitem{Wishart1928} Wishart J., \emph{The Generalized Product Moment Distribution in Samples from a Normal Multivariate Population}, Biometrika \textbf{A 20} (1928) 32.

\bibitem{VoiculescuDykemaNica1992} Voiculescu D. V., Dykema K. J., Nica A., \emph{Free Random Variables}, CRM Monograph Series, Vol. 1, Am. Math. Soc., Providence, 1992.

\bibitem{Speicher1994} Speicher R., \emph{Multiplicative functions on the lattice of non--crossing partitions and free convolution}, Math. Ann. \textbf{298} (1994) 611.

\bibitem{BurdaJaroszJurkiewiczNowakPappZahed2009} Burda Z., Jarosz A., Jurkiewicz J., Nowak M. A., Papp G., Zahed I., \emph{Applying Free Random Variables to Random Matrix Analysis of Financial Data}, Quantitative Finance (2011)\textbf{11} (7), 1103--114 .

\bibitem{BurdaJaroszNowakSnarska2010}
{\sc Burda, Z., Jarosz, A., Nowak, M., and Snarska, M.}
\newblock A random matrix approach to varma processes.
\newblock {\em New Journal of Physics 12\/} (2010), 075036.
\newblock \href{http://arxiv.org/abs/1002.0934}{[arXiv:1002.0934v1]}.


\bibitem{SilversteinBai1995}
{\sc Silverstein, J., and Bai, Z.}
\newblock On the empirical distribution of eigenvalues of a class of large
  dimensional random matrices.
\newblock {\em Journal of Multivariate Analysis 54}, 2 (1995), 175 -- 192.
\newblock
  \href{http://citeseerx.ist.psu.edu/viewdoc/summary?doi=10.1.1.40.4138}{[DOI:10.1006/jmva.1995.1051]}.


\bibitem{MarcenkoPastur1967}
{\sc Marcenko, V., and Pastur, L.}
\newblock Distribution of eigenvalues for some sets of random matrices.
\newblock {\em Math. USSRSbornik 1}, 4 (1967), 457--483.
\newblock
  \href{http://iopscience.iop.org/0025-5734/1/4/A01}{[DOI:10.1070/SM1967v001n04ABEH001994]}.

\bibitem{PafkaPottersKondor2004}
{\sc Pafka, S., Potters, M., and Kondor, I.}
\newblock Exponential weighting and random--matrix--theory--based filtering of
  financial covariance matrices for portfolio optimization.
\newblock
  \href{http://arxiv.org/abs/cond-mat/0402573}{[arXiv:cond-mat/0402573]}, 2004.

\bibitem{ThurnerBiely2007}
{\sc Thurner, S., and Biely, C.}
\newblock The eigenvalue spectrum of lagged correlation matrices.
\newblock {\em Acta Physica Polonica \textbf{B} 38\/} (2007), 4111.
\newblock \href{http://arxiv.org/abs/physics/0609053}{[arXiv:physics/0609053]}.

\bibitem{BouchaudLalouxAugustaMiceliPotters2007}
{\sc Bouchaud, J., Laloux, L., Miceli, M., and Potters, M.}
\newblock Large dimension forecasting models and random singular value spectra.
\newblock {\em The European Physical Journal B - Condensed Matter and Complex
  Systems 55}, 2 (2007), 201 -- 207.
\newblock
  \href{http://arxiv.org/abs/physics/0512090}{[arXiv:physics/0512090v1]}.

\bibitem{EdelmanRao2005}
{\sc Edelman, A., and Raj~Rao, N.}
\newblock Random matrix theory.
\newblock {\em Acta Numerica 14\/} (2005), 233--297.
\newblock \href{http://www.eecs.umich.edu/~rajnrao/Acta05rmt.pdf}{[e-print]}.

\bibitem{Snarska2008}
{\sc Snarska, M.}
\newblock Toy model for large non-symmetric random matrices.
\newblock {\em Acta Physica Polonica A 114}, 3 (2008).
\newblock \href{http://arxiv.org/abs/1004.4522}{[arXiv:1004.4522]}.


\bibitem{Granger2001}
{\sc Granger, C.}
\newblock Macroeconometrics - past and future.
\newblock {\em Journal of Econometrics 100}, 1 (2001), 17 -- 19.
\newblock
  \href{http://cenet3.nsd.edu.cn/upload/5.33.51-01-7-7-9-11-7.pdf}{[DOI:10.1016/S0304-4076(00)00047-6]}.


\bibitem{BollerslevEngleWooldridge1998}
{\sc Bollerslev, T., Engle, R., and Wooldridge, J.}
\newblock A capital asset pricing model with time-varying covariances.
\newblock {\em The Journal of Political Economy 96\/} (1988), 116--131.
\newblock \href{http://harrisd.net/papers/ARCHSV/Multivariate
  ARCH/BollerslevEngleWooldridge1988JoPE.pdf}{[JSTOR:1830713]}.

\bibitem{Geweke1997}
{\sc Geweke, J.}
\newblock The dynamic factor analysis of economic time series.
\newblock In {\em Latent Variables in Social Economic Models}, D.~Aigner and
  A.~Goldberger, Eds. North Holland, Amsterdam, 1997.

\bibitem{StockWatson2002-1}
{\sc Stock, J., and Watson, M.}
\newblock Macroeconomic forecasting using diffusion indexes.
\newblock {\em Journal of Business and Economic Statistics 20\/} (2002),
  147--162.
\newblock
  \href{http://www.eui.eu/Personal/Banerjee/courses/winter2004/SW2002.pdf}{[DOI:10.1198/073500102317351921]}.

\bibitem{StockWatson2002-2}
{\sc Stock, J., and Watson, M.}
\newblock Forecasting using principal components from a large number of
  predictors.
\newblock {\em Journal of the American Statistical Association 97\/} (2002),
  1167--1179.
\newblock \href{http://www.jstor.org/pss/3085839}{[JSTOR:3085839]}.

\bibitem{StockWatson2005}
{\sc Stock, J., and Watson, M.}
\newblock Implications of dynamical factor models for var analysis.
\newblock working paper
  \href{http://www.princeton.edu/~mwatson/papers/favar.pdf}{[NBER-eprint]},
  2005.

\bibitem{ForniHallinLippiRechlin2000}
{\sc Forni, M., Hallin, M., Lippi, M., and Reichlin, R.}
\newblock The generalized dynamic factor model:identification and estimation.
\newblock {\em The Review of Economic and Statistics 82\/} (2000), 540--554.
\newblock
  \href{http://www.eabcn.org/research/documents/rev3.pdf}{[DOI:10.1162/003465300559037]}.

\bibitem{ForniHallinLippiRechlin2004}
{\sc Forni, M., Hallin, M., Lippi, M., and Reichlin, R.}
\newblock The generalized dynamic factor model: Consistency and rates.
\newblock {\em Journal of Econometrics 119\/} (2004), 231--255.
\newblock
  \href{http://www.economia.unimore.it/forni_mario/consrates.pdf}{[DOI:10.1016/S0304-4076(03)00196-9]}.

\bibitem{Bai2003}
{\sc Bai, J.}
\newblock Inferential theory for factor models of large dimensions.
\newblock {\em Econometrica 71\/} (2003), 135--171.
\newblock
  \href{http://www.econ.nyu.edu/user/baij/asymfacR2.pdf}{[DOI:10.1111/1468-0262.00392]}.

\bibitem{BaiNg2002}
{\sc Bai, J., and Ng, S.}
\newblock Determining the number of factors in approximate factor model.
\newblock {\em Econometrica 70\/} (2002), 191--221.
\newblock
  \href{http://www.econ.nyu.edu/user/baij/econometrica02.pdf}{[DOI:10.1111/1468-0262.00273]}.

\bibitem{LalouxCizeauBouchaudPotters1999} Laloux L., Cizeau P., Bouchaud J.--P., Potters M., \emph{Noise Dressing of Financial Correlation Matrices}, Phys. Rev. Lett. \textbf{83} (1999) 1467 [\texttt{arXiv:cond-mat/9810255}].

\bibitem{BurdaJurkiewicz2004} Burda Z., Jurkiewicz J., \emph{Signal and Noise in Financial Correlation Matrices}, Physica \textbf{A 344} (2004) 67 [\texttt{arXiv:cond-mat/0312496}].

\bibitem{BurdaGorlichJaroszJurkiewicz2004} Burda Z., G\"{o}rlich A., Jarosz A., Jurkiewicz J., \emph{Signal and Noise in Correlation Matrix}, Physica \textbf{A 343} (2004) 295 [\texttt{arXiv:cond-mat/0305627}].

\bibitem{FriedbergInselSpence2002}
{\sc Friedberg, S., Insel, A., and Spence, L.}
\newblock {\em Linear Algebra}.
\newblock Prentice Hall, 2002.


\bibitem{Voiculescu1991}
{\sc Voiculescu, D.~V.}
\newblock Limit laws for random matrices and free products.
\newblock {\em Invent. Math. 104\/} (1991), 201.
\newblock
  \href{http://www.kryakin.com/files/Invent_mat_(2_8)/104/104_07.pdf}{[DOI:10.1007/BF01245072]}.



\bibitem{BernankeBoivin2003}
{\sc Bernanke, B., and Boivin, J.}
\newblock Monetary policy in a data rich environment.
\newblock {\em Journal of Monetary Economics 50\/} (2003), 525.
\newblock
  \href{http://faculty.wcas.northwestern.edu/~lchrist/finc520/sdarticle.pdf}{[DOI:10.1016/S0304-3932(03)00024-2]}.

\bibitem{StockWatson1999}
{\sc Stock, J., and Watson, M.}
\newblock Forecasting inflation.
\newblock {\em Journal of Monetary Economics 44\/} (1999), 293--335.
\newblock
  \href{http://homes.chass.utoronto.ca/~melino/eco327/stock.pdf}{[SSRN:155850]}.

%
%
%
%
%
%
%
%
%
%


\end{thebibliography}
\end{document}